# Modeling transport and mean age of dense core vesicles in large axonal arbors

I. A. Kuznetsov[(a), (b)] and A. V. Kuznetsov[(c)]

[(a)] Perelman School of Medicine, University of Pennsylvania, Philadelphia, PA 19104, USA

[(b)] Department of Bioengineering, University of Pennsylvania, Philadelphia, PA 19104, USA

[(c)] Department of Mechanical and Aerospace Engineering, North Carolina State University, Raleigh, NC 27695-7910, USA; e-mail: avkuznet@ncsu.edu

**Abstract**

A model simulating transport of dense core vesicles (DCVs) in type II axonal terminals of *Drosophila* motoneurons has been developed. The morphology of type II terminals is characterized by the large number of *en passant* boutons. The lack of both scaled up DCV transport and scaled down DCV capture in boutons results in a less efficient supply of DCVs to distal boutons. Furthermore, the large number of boutons that DCVs pass as they move anterogradely, until they reach the most distal bouton, may lead to the capture of a majority of DCVs before they turn around in the most distal bouton to move in the retrograde direction. This may lead to a reduced retrograde flux of DCVs and a lack of DCV circulation in type II terminals. The developed model simulates DCV concentrations in boutons, DCV fluxes between the boutons, age density distributions of DCVs, and the mean age of DCVs in various boutons. Unlike published experimental observations, our model predicts DCV circulation in type II terminals after these terminals are filled to saturation. This disagreement is likely because experimentally observed terminals were not at steady-state, but rather were accumulating DCVs for later release. Our estimates show that the number of DCVs in the transiting state is much smaller than that in the resident state. DCVs traveling in the axon, rather than DCVs transiting in the terminal, may provide a reserve of DCVs for replenishing boutons after a release. The techniques for modeling transport of DCVs developed in our paper can be used to model the transport of other organelles in axons.

# 1. Introduction

Investigating axons with large arbors is important for understanding Parkinson's disease (PD) [1]. Loss of dopaminergic neurons in the substantia nigra is implicated in PD pathogenesis. The extensive axonal arborization of these neurons may put them at an increased risk, possibly due to the resulting tight energy budget [2] and the potential inability to supply organelles to synaptic boutons by a single axon sufficiently, especially to distally located boutons [3,4].

In order to understand transport of organelles in large arbors, simpler animal models can be examined. *Drosophila melanogaster* is a popular model for basic studies of the nervous system [5]. Varicosities located along the axon terminals are called *en passant* boutons (hereafter referred to as boutons); they are active terminal sites that release neurotransmitters. Three types of boutons with different morphologies are identified in *Drosophila*: I, II, and III. Type I boutons are further divided into Ib and Is boutons, which represent big and small boutons, respectively [5-7].

Neuropeptides, a type of signal molecules that transmit signals across the synaptic cleft, are synthesized in the soma and transported toward the synapses in the dense core vesicles (DCVs) by means of fast axonal transport [8,9]. Ref. [3] investigated DCV transport in axons with type II endings, which are characterized by the greatest number of boutons when compared to the other *Drosophila* motoneurons. Here, we investigate how the issue of potentially insufficient supply with organelles of distal boutons in large axonal arbors is solved in motoneurons with type II endings [10].

Ref. [11] suggested that neuropeptide delivery to boutons is based on the sporadic capture of DCVs from the circulation that these authors discovered in type Ib terminals. However, ref. [12] reported that large neuropeptide content in type III boutons is due to increased DCV capture rather than delivery, which results in fewer transiting DCVs and a lack of DCV circulation in type III terminals. Furthermore, ref. [3] reported an inefficient supply of DCVs to distal boutons and a lack of DCV circulation in type II terminals. In this paper, we attempt to better understand the lack of similarity between type Ib and type II terminals.

The model simulating DCV transport in type Ib and type III axon terminals was developed in refs. [13-15]. Ref. [16] investigated the effects of reversibility of DCV capture in boutons. In the present paper, we extend our model to simulate DCV transport in type II terminals, which contain a much larger number of boutons. Additionally, our model is now capable of evaluating the mean age of DCVs and their age density in various boutons. We compared the mean ages of DCVs in proximal and distal boutons as predicted by the model with those observed experimentally and reported in ref. [3]. Our DCV age simulations assumed that after being captured into a resident state, the captured DCVs are eventually destroyed in a bouton. The lack of agreement with the experimental results of ref. [3] suggests that DCVs captured into the resident state in boutons are not destroyed in boutons, but rather are re-released from the resident state back to the transiting pool.

## 2. Materials and models

### 2.1. Governing equations

We assumed that DCVs are synthesized in the soma, transported in the axon, and then enter one of the terminal branches where they can be captured in boutons (Fig. 1). We assumed that an axon has three identical branches (Fig. 1a). We followed ref. [11] and numbered the boutons #1 to #26 from the most distal to the most proximal (Fig. 1a).

There are two pools of DCVs in this system [11]: resident DCVs in boutons and transiting DCVs that travel in the axon and terminal (Table 1). We assume that the number of resident DCVs in the whole region containing boutons greatly exceeds the number of transiting DCVs. This is supported by the following estimate. The number of DCVs in the transiting state can be estimated as $j_{ax->26}(0) L_{26} / v$, where $j_{ax->26}(0)$ is the flux from the axon to the most proximal bouton at $t = 0$, $L_{26}$ is the length of a compartment occupied by the most proximal bouton (10 μm), and $v$ is the average DCV velocity, which is estimated to be ~1 μm/s [11,17-19]. This estimates that the number of DCVs transiting in the terminal is ~0.33 vesicles, which is about one percent of the number of vesicles in the resident state (~34 DCVs reside in the most proximal type II bouton, [3]).

The variables utilized in our model are summarized in Table 2.

Table 1. Various pools of DCVs in the axon and terminal.

| Type of DCVs | Location | Possible fate(s) |
|---|---|---|
| Resident | Boutons | Resident DCVs can re-enter the transiting pool (this scenario is simulated by $\delta = 1$) or be destroyed in boutons (simulated by $\delta = 0$). If $0 < \delta < 1$, a portion of DCVs re-enter the transiting pool after spending |

| | | some time in boutons and some are destroyed in boutons. |
| --- | --- | --- |
| Transiting | Axon and terminal | The axon contains only transiting DCVs. In a terminal, transiting DCVs can travel from bouton to bouton, either anterogradely or retrogradely, and can be captured into the resident state in boutons. The capture rate is determined by the capture efficiency of boutons $\varpi$. For $\delta = 1$ the capture is reversible, and after spending some time in the resident state in boutons (determined by the half-residence time of DCVs in boutons, $T_{1/2}$) the captured DCVs return back to the transiting pool. For $\delta = 0$ the capture is irreversible, and after spending some time in boutons (determined by the half-life of DCVs in boutons, also denoted as $T_{1/2}$) the DCVs are destroyed. |

Table 2. Model variables.

| Symbol | Definition | Units |
| --- | --- | --- |
| $j_{a \to b}(t)$ | Flux of DCVs from compartment "a" to compartment "b" (see Fig. 1b) | vesicles/s |
| $n_{ax}(t)$ | Average concentration of DCVs in the axon (axon is modeled as a single compartment) | vesicles/μm |
| $n_i(t)$ | Concentrations of resident DCVs in bouton $i$ ($i=1,\ldots,26$) | vesicles/μm |
| $t$ | Time | s |

As DCV transport in an axon is essentially one-dimensional, we characterized the DCV concentration by its linear number density, which we defined as the number of DCVs per unit length of the axon. The concentration of DCVs residing in a bouton can change either due to DCV capture into the resident state (as DCVs pass the bouton) or by DCV release back to the transiting state (alternatively by DCV destruction in boutons). DCV capture is shown in Fig. 1b by block arrows. We used a multi-compartment model [20-22] to develop the governing equations in 27 compartments: the axon and 26 boutons (Fig. 1). The number of DCVs in a compartment is the conserved property.

We first state the conservation of DCVs in the resident state in the most proximal bouton:

$$L_{26} \frac{dn_{26}}{dt} = \min\left[h_{26}^a \left(n_{sat0,26} - n_{26}\right), j_{ax \to 26}\right] + \min\left[h_{26}^r \left(n_{sat0,26} - n_{26}\right), j_{25 \to 26}\right] - L_{26} \frac{n_{26} \ln(2)}{T_{1/2}}. \quad (1)$$

The meaning of different terms in Eqs. (1)-(3) is explained in Table 3.

Table 3. Significance of various terms in the conservation equations for DCVs in the resident state in boutons. Applies to Eqs. (1)-(3).

| Physical process | Terms that simulate the process | Notes |
|---|---|---|
| Accumulation of transiting DCVs into the resident state | The term on the left-hand side of Eqs. (1)-(3) | |
| DCV capture into the resident state as DCVs pass the bouton anterogradely or retrogradely | The first and second terms on the right-hand side of Eqs. (1) and (2). In Eq. (3) DCV capture is described by a single term only (the first term on the right-hand side) because DCVs pass bouton 1 only once. | If the number of DCVs in the transiting state is sufficiently large, the rate of DCV capture by a bouton is assumed to be proportional to $n_{sat0,i} - n_i$ (the difference between the steady-state DCV concentration in a bouton for infinite DCV half-residence time, or infinite half-life, and the current DCV concentration in the bouton). The DCV concentration for infinite DCV half-residence time, or infinite half-life, is used to allow DCV capture to continue even at steady-state if $T_{1/2}$ is finite. This is necessary to compensate for the DCV re-release from boutons or their destruction. The conservation equations are not affected by how DCVs leave the resident state, whether they are released back to the transiting pool or destroyed in boutons. The MIN function simulates the fact that the rate of DCV capture into the resident state cannot exceed the flux of DCV into the bouton. Note that retrograde fluxes between the boutons will initiate with a 300 s delay (see Eqs. (8)-10)) that is |

| | | required for DCVs to change anterograde to retrograde motors in bouton 1 [23]. Hence, the capture of retrogradely moving DCVs, described by the second term on the right-hand side of Eqs. (1) and (2), will initiate 300 s after the DCVs start entering the terminal. |
|---|---|---|
| Release back into the transiting state (or destruction in the bouton) | The last term on the right-hand side of Eqs. (1)-(3) | |

The conservation of DCVs in the resident state in boutons 25 through 2 is stated as follows:

$$L_i \frac{dn_i}{dt} = \min\left[h_i^a \left(n_{sat0,i} - n_i\right), j_{i+1 \to i}\right] + \min\left[h_i^r \left(n_{sat0,i} - n_i\right), j_{i-1 \to i}\right] - L_i \frac{n_i \ln(2)}{T_{1/2}} \quad (i=25,24,\ldots,2).$$

(2)

In the most distal bouton (bouton 1), the requirement of DCV conservation in the resident state produces the following equation:

$$L_1 \frac{dn_1}{dt} = \min\left[h_1 \left(n_{sat0,1} - n_1\right), j_{2 \to 1}\right] - L_1 \frac{n_1 \ln(2)}{T_{1/2}}.$$

(3)

In addition to the conservation of resident DCVs in boutons, a statement can be made based on conservation of DCVs traveling in the axon, which results in the following equation:

$$L_{ax} \frac{dn_{ax}}{dt} = j_{soma \to ax} + 3 j_{26 \to ax} - 3 j_{ax \to 26} - L_{ax} \frac{n_{ax} \ln(2)}{T_{1/2,ax}}.$$

(4a)

Our model does not simulate DCV processing in the soma (processes such as the production of DCVs and destruction of returning DCVs in the somatic lysosomes). Therefore, $j_{soma \to ax}$ in Eq. (4a) should be interpreted as the net DCV flux from the soma into the axon, calculated as the DCV production rate in the soma minus the DCV destruction rate in the somatic lysosomes.

We also consider the situation when the flux of DCVs from the axon to the most proximal bouton remains constant. In this case, Eq. (4a) must be replaced with the following equation:

$$\frac{dn_{ax}}{dt} = 0.$$

(4b)

Eqs. (1)-(4) include DCV fluxes from the axon to the most proximal bouton, $j_{ax \to 26}$, and back to the axon, $j_{26 \to ax}$, as well as anterograde and retrograde fluxes between the boutons (Fig. 1b). These fluxes need to be modeled. We assumed that the DCV flux from the axon to the most distal bouton is proportional to the average DCV concentration in the axon:

$$j_{ax \to 26} = h_{ax} n_{ax}. \tag{5}$$

Here we follow our previous paper [16] where we developed the method of simulating the effect of re-entry of resident DCVs into the transiting pool. The portion of DCVs that escape from the captured state in boutons back into the transiting pool was characterized by parameter $\delta$. The case when all captured DCVs eventually re-enter the transiting pool is simulated by $\delta = 1$ and the case when all captured DCVs are eventually destroyed in boutons is simulated by $\delta = 0$. The effect of adopting one of these hypotheses should be carefully analyzed. Indeed, the hypothesis that DCVs captured into the resident state in boutons are eventually re-released to the transiting pool ($\delta = 1$) is supported by the fact that organelles like DCVs are usually destroyed in lysosomes, which are abundant in the soma but not in the terminals. On the other hand, the hypothesis that the released DCVs return to the soma for degradation requires a non-zero retrograde flux from the terminal back to the axon, which seemingly contradicts Tao et al. [3] who reported almost no retrograde flux in type II terminals.

Equations for the DCV fluxes between the boutons, $j_{26 \to 25}, \ldots, j_{2 \to 1}$ (Fig. 1b), are written by stating the conservation of DCVs in the transiting state. The anterograde flux between the most proximal bouton (bouton 26) and bouton 25 is

$$j_{26 \to 25} = j_{ax \to 26} - \min\left[ h_{26}^a \left( n_{sat0,26} - n_{26} \right), j_{ax \to 26} \right] + \delta \varepsilon L_{26} \frac{n_{26} \ln(2)}{T_{1/2}}. \tag{6}$$

The meaning of different terms on the right-hand side of Eqs. (6)-(10) is explained in Table 4.

Table 4. Significance of various terms in the equations for DCV fluxes between the boutons. Applies to Eqs. (6)-(10).

| Physical process | Terms that simulate the process | Notes |
|---|---|---|
| Difference between the flux of DCVs entering a bouton and the DCV capture rate | The first and second terms on the right-hand side of Eqs. (6)-(10) | |

| into the resident state in the bouton | | |
|---|---|---|
| Rate of DCV release from the resident state back to transiting state | The last term on the right-hand side of Eqs. (6)-(10) | If $\delta=0$, then all DCVs are destroyed in boutons and none are released back to the transiting pool. On the other hand, if $\delta=1$, then all DCVs are released to the transiting pool after spending some time in the resident state. Parameter $\varepsilon$ describes how the DCVs that are released from the transiting state are split between the anterogradely and retrogradely moving pools ($\varepsilon$ is the portion of DCVs that join the anterograde pool and $(1-\varepsilon)$ is the portion that join the retrograde pool). The multiplier $H[t-t_1]$ in Eqs. (8)-(10) accounts for the 300 s delay that it takes for the retrograde flux to start. |

Anterograde fluxes between boutons 25 through 2 are modeled by the following equations:

$$j_{i \to i-1} = j_{i+1 \to i} - \min\left[h_i^a\left(n_{sat0,i} - n_i\right), j_{i+1 \to i}\right] + \delta \varepsilon L_i \frac{n_i \ln(2)}{T_{1/2}} \qquad (i=25,24,\ldots,2). \qquad (7)$$

The retrograde flux from bouton 1 into bouton 2 is

$$j_{1 \to 2} = H[t-t_1]\left\{j_{2 \to 1} - \min\left[h_1\left(n_{sat0,1} - n_1\right), j_{2 \to 1}\right]\right\} + \delta L_1 \frac{n_1 \ln(2)}{T_{1/2}}, \qquad (8)$$

where $H$ is the Heaviside step function.

Note that the last term in Eq. (8) does not contain $\varepsilon$. This is because DCVs released from the resident state in this bouton can only join the pool of retrogradely moving vesicles.

Retrograde fluxes between boutons 2 through 25 are modeled by the following equations:

$$j_{i \to i+1} = H[t-t_1]\left\{j_{i-1 \to i} - \min\left[h_i^r\left(n_{sat0,i} - n_i\right), j_{i-1 \to i}\right]\right\} + \delta(1-\varepsilon) L_i \frac{n_i \ln(2)}{T_{1/2}} \qquad (i=2,3,\ldots,25).$$

$$(9)$$

Finally, the retrograde flux from bouton 26 back to the axon is modeled by the following equation:

$$j_{26 \to ax} = H[t - t_1]\{j_{25 \to 26} - \min[h_{26}^r(n_{sat0,26} - n_{26}), j_{25 \to 26}]\} + \delta(1 - \varepsilon)L_{26}\frac{n_{26}\ln(2)}{T_{1/2}}. \tag{10}$$

In Eqs. (5)-(10) fluxes have units of vesicles/s. Eqs. (1)-(4) describe a system of 27 first-order ordinary differential equations that requires 27 initial conditions. In *Drosophila* release of DCVs by exocytosis, which is associated with molting behavior, can be massive, as much as ~90% of content [23]. Since is it not our goal to simulate the specific experiments reported in ref. [3] but rather to gain a fundamental understanding of DCV transport in large axonal arbors, for simplicity we simulate the process of refilling the terminal after a compete release, which means that initially the terminal is empty:

$$n_1(0) = 0, \ldots, n_{26}(0) = 0, \ n_{ax}(0) = n_{sat,ax}. \tag{11}$$

The last equation in (11) assumes that initially the axon is filled to saturation. Since the terminal is initially empty, DCVs enter the terminal at a large rate earlier on in the simulation. If the rate of DCV synthesis in the soma is small, there may be an initial decrease in the concentration of transiting DCVs in the axon. The axonal DCV concentration will recover when the boutons in the terminal are filled to (or close to) saturation, and thus less DCVs are needed in the terminal.

**2.2. Estimation of values of parameters involved in the model**

We used two methods to estimate the values of model parameters. First, we estimated the parameters whose values we could find in published literature or assume on physical grounds. These are $j_{ax \to 26}$, $L_1, \ldots, L_{26}$, $L_{ax}$, $t_1$, $T_{1/2}$, $T_{1/2,ax}$, and $\varpi$. We summarized the values of these parameters in Table 5. We then estimated the values of parameters which we were unable to find in the literature ($n_{sat,ax}$, $n_{sat,1}, \ldots, n_{sat,26}$, $h_1, \ldots, h_{26}$, $h_{in}$, $n_{sat0,1}, \ldots, n_{sat0,26}$, $n_{sat,ax}$, and $j_{soma}$) by stating DCV conservation at the initial moment ($t = 0$) and at steady-state. We reported the values of these parameters in Table 6. (In what follows, the words "saturated" and "steady-state" are used interchangeably.)

Table 5. Model parameters estimated based on values found in the literature or assumed on physical grounds.

| Symbol | Definition | Units | Estimated value(s) or | Reference(s) |
|--------|------------|-------|----------------------|--------------|

| | | | range | |
|---|---|---|---|---|
| $a$ | Parameter characterizing a decrease in the number of DCVs in the saturated state from the most proximal to the most distal bouton, as defined in Eq. (12) | | 1.1 [a] | |
| $j_{ax \to 26}$ | Flux from the axon to the most proximal bouton | vesicles/s | 0.0333 [b] | [3,11] |
| $L_1,\ldots,L_{26}$ | Lengths of compartments occupied by boutons 1, 2,…, 26 (defined in Fig. 1a) | μm | 10 [c] | [24] |
| $L_{ax}$ | Length of the axon, see Fig. 1a | μm | 500 | [24] |
| $t_1$ | Time required for DCVs to change the direction in the most distal bouton, if they are not captured | s | 300 | [11,23] |
| $T_{1/2}$ | Half-life or half-residence time of captured DCVs | s | $2.16 \times 10^4$ | [25] |
| $T_{1/2,ax}$ | Half-life of DCVs in the axon | s | $(1\ldots100) \times T_{1/2}$ [d] | [23,25] |
| $\varepsilon$ | Parameter simulating how DCVs released from the resident state in boutons are split between the anterograde and retrograde transiting pools. $\varepsilon$ is the portion of released DCVs that join the anterograde pool while $(1-\varepsilon)$ is the portion of released DCVs that join the retrograde pool. | | 0.5 [e] | |
| $\delta$ | Parameter determining the fate of DCVs captured into the resident state in boutons. $\delta = 0$ simulates the situation when all DCVs are eventually destroyed in boutons while $\delta = 1$ simulates the situation when DCVs, after spending some time in the resident state, are released back to the transiting pool. | | 0-1 | |
| $\varpi$ | Capture efficiency defined as the percentage of DCVs captured in a bouton when DCVs pass the bouton. In boutons 2,…,26 | | 0.1 [f] | [3] |

| | capture occurs twice, when DCVs pass the bouton in anterograde and in retrograde directions. | | | |
|---|---|---|---|---|

[a] Ref. [3] reported a decrease in the DCV concentration in distal boutons compared to proximal boutons in type II terminals.

[b] According to ref. [3], the anterograde flux into a terminal with type II boutons is half compared to that into a terminal with type Ib boutons (~ 4 vesicles/min, [11]), that is ~2 vesicles/min.

[c] The length of a compartment occupied by a bouton equals the spacing between two adjacent boutons (Fig. 1), which is ~10 μm according to ref. [24].

[d] DCVs are transported over large distances in the axons (up to 1 m in humans). This means that they must somehow be protected from degradation in the axon, and thus the DCV half-life in the axon is probably larger than the DCV half-life or half-residence time in boutons. A possible physical mechanism explaining such protection is the scarcity of organelle degradation machinery in axons [23]. We investigated the effect of DCV half-life in the axon on DCV transport, specifically cases when $T_{1/2,ax}$ varies in the range between ($T_{1/2}, 100 \times T_{1/2}$).

[e] There is no experimental data that indicate how the DCVs are split. For computations presented in this paper we assumed that $\varepsilon = 0.5$. In future research, a sensitivity analysis similar to that reported in refs. [26,27], with respect to parameter $\varepsilon$, as well as to other parameters, should be performed.

[f] For type II boutons, we assumed that $\varpi = 0.1$, the same value when DCVs pass the bouton in anterograde or retrograde directions, for all 26 boutons. This estimate is based on data reported in ref. [3]. Data presented in Fig. 6E of ref. [3] may suggest that retrograde capture is slightly greater. However, since in a terminal with type II boutons the retrograde DCV flux is small compared to the anterograde flux, this difference would not affect our model.

### 2.2.1. Saturated DCV concentrations in boutons, $n_{sat,i}$ ($i = 1,…,26$)

Type II terminals, studied in ref. [3], have ~80 boutons per muscle, usually distributed on 3 or 4 branches [24]. We thus estimated that there are ~26 boutons per single branch. Based on ref. [3], we estimated that in the saturated state, there are ~34 DCVs in the most proximal type II bouton.

Also, ref. [3] reported a decreased DCV content in distal boutons; for example, neuropeptide release was 50% lower for distal boutons compared to proximal boutons. To model this, we assumed that the number of DCVs in the saturated state decreases in more distal boutons as $34/a^{26-i}$, where $a=1.1$ and $i$ is the number of a bouton (Fig. 1). Since spacing between boutons is $L_i \sim 10$ µm [24], the saturated concentration of DCVs in type II boutons is

$$n_{sat,i} = (34/L_i)/a^{26-i} = 3.4/a^{26-i} \text{ vesicles/µm } (i = 1,\ldots,26). \tag{12}$$

Eq. (12) postulates a set capacity for vesicles (e.g. akin to parking spaces) that limits accumulation. This allows boutons with excess supply of vesicles to fill to a set amount and allow more vesicles to continue traveling distally.

One of the goals of our research is to estimate whether it is realistic to assume that a saturated state can be reached in type II terminals. The reason why it may not be reached is because in the process of transitioning from being a larva to a fly, *Drosophila* larva retracts the neurons, destroys the muscles, builds new muscles, and reinnervates [24]. This may happen before transport processes in type II terminals reach steady-state.

### 2.2.2. Average saturated DCV concentration in the axon, $n_{sat,ax}$

We assumed, following ref. [14], that the average DCV concentration in the axon is 10% of the saturated DCV concentrations in boutons. This leads to the following estimate:

$$n_{sat,ax} = 0.1 \times 3.4 \text{ vesicles/µm} = 0.34 \text{ vesicles/µm}. \tag{13}$$

### 2.2.3. Mass transfer coefficient characterizing the rate at which DCVs enter the most proximal bouton from the axon, $h_{ax}$

We assumed that the flux from the axon into the terminal branch is proportional to the average DCV concentration in the axon:

$$j_{ax \to 26} = h_{ax} n_{sat,ax}. \tag{14}$$

Since the flux into a type II terminal branch is ~2 vesicles/min (see footnote "b" after Table 2), by solving Eq. (14) for $h_{ax}$ we obtained that

$$h_{ax} = 0.0980 \text{ µm/s}. \tag{15}$$

**2.2.4. Mass transfer coefficients characterizing DCV capture into the resident state in boutons, $h_1$, $h_2$,..., $h_{26}$; and saturated concentrations of DCVs in boutons at infinite DCV half-life or half-residence time, $n_{sat0,1}$, $n_{sat0,2}$,..., $n_{sat0,26}$**

We assumed that the mass transfer coefficients characterizing DCV capture when DCVs pass a bouton in anterograde and retrograde directions are equal:

$$h_i^a = h_i^r = h_i \qquad (i=2,\ldots,26). \tag{16}$$

We also assumed that initially, there are no resident DCVs in the boutons. We then wrote equations simulating the reduction of the DCV flux after DCVs pass boutons 26, 25,..., and 1, respectively, at $t = 0$. Capture efficiency, $\varpi$, characterizes capture initiation [3]. Therefore, after passing each bouton, the DCV flux is initially reduced by 10%, and we can state the following:

$$\varpi \, j_{ax \to 26} = h_{26} \left( n_{sat0,26} - 0 \right), \tag{17a}$$

...

$$\varpi \, j_{ax \to 26} (1-\varpi)^{26-i} = h_i \left( n_{sat0,i} - 0 \right), \tag{17b}$$

...

$$\varpi \, j_{ax \to 26} (1-\varpi)^{25} = h_1 \left( n_{sat0,1} - 0 \right), \tag{17c}$$

On the other hand, at steady-state, the rate at which DCVs are captured must be equal to the rate at which they are destroyed (or re-enter the transiting pool). It should be noted that steady-state may never be reached in type II terminals in *Drosophila* because larva may transition into a fly before it is reached, but the model must still be able to simulate the steady-state situation. This leads to the following equation:

$$2 h_i \left( n_{sat0,i} - n_{sat,i} \right) = L_i \frac{n_{sat,i} \ln(2)}{T_{1/2}} \qquad (i = 2,\ldots,26). \tag{18}$$

A factor of two on the left-hand side of Eq. (18) appears because DCVs have two chances to be captured in boutons 2,...26: they can be captured as they travel anterogradely and retrogradely through a bouton.

In a similar equation describing DCV conservation at steady-state in bouton 1, the factor of two on the left-hand side is absent because DCVs pass this bouton only once:

$$h_1\left(n_{sat0,1} - n_{sat,1}\right) = L_1 \frac{n_{sat,1} \ln(2)}{T_{1/2}}. \tag{19}$$

Eqs. (17)-(19) were solved by using Matlab's (Matlab R2018b, MathWorks, Natick, MA, USA) SOLVE solver, and the results are summarized in Tables S1 and S2.

**2.3. The net rate of DCV production in the soma**

Following ref. [24], we assumed that an axon splits into three branches. Ref. [11] argued that there is no active address system for directing DCVs to a particular bouton. Also, the results reported in ref. [3] suggest the depletion of distal boutons of DCVs, which would not be likely if an actively controlled DCV delivery system existed. Therefore, we assumed that the DCV transport is passively regulated and neglected any possible feedback effects on DCV transport.

At steady-state, the net rate of DCV production in the soma, $j_{soma}$ (defined as the rate of DCV synthesis minus the rate of DCV destruction in somatic lysosomes), must be equal to the rate of DCV destruction in the axon and three branches:

$$j_{soma} = 3\sum_{i=1}^{26} L_i n_{sat,i} (1-\delta) \frac{\ln(2)}{T_{1/2}} + L_{ax} n_{sat,ax} \frac{\ln(2)}{T_{1/2,ax}}. \tag{20}$$

According to our model, the DCV flux from the soma into the axon is heavily dependent on the fate of DCVs in boutons. If $\delta = 0$, then all DCVs captured in boutons are eventually destroyed. To maintain steady-state, this scenario requires more DCVs to enter the axon than the scenario in which DCVs captured in boutons are re-released and reenter the transiting pool ($\delta = 1$).

Table 6. Model parameters estimated based on DCV balances at the initial state and at steady-state as well as on values reported in Table 5 (see section 2.2 for the details on how the estimates were done).

| Symbol | Definition | Units | Estimated value(s) |
|---|---|---|---|
| $h_1$, $h_2$,…, $h_{26}$ | Mass transfer coefficients characterizing the rates of capture of DCVs into the | μm/s | Data are summarized in Table S1 |

| | resident state in boutons 1,…,26, respectively (for boutons 2,…,26 we assumed that $h_i^a = h_i^r = h_i$) | | |
|---|---|---|---|
| $h_{ax}$ | Mass transfer coefficient characterizing the rate at which DCVs leave the axon and enter the most proximal bouton | µm/s | 0.0980 |
| $j_{soma}$ | The rate of DCV synthesis in the soma minus the rate of DCV destruction in the somatic lysosomes | vesicles/s | Calculated by Eq. (20) |
| $n_{sat,1},…,n_{sat,26}$ | Saturated (steady-state) concentrations of DCVs in boutons 1,…,26, defined by Eq. (12) | vesicles/µm | $3.4/a^{26-i}$ |
| $n_{sat0,1},…,n_{sat0,26}$ | Saturated concentrations of DCVs in boutons 1,…,26, respectively, at infinite DCV half-life or at infinite DCV residence time | vesicles/µm | Data are summarized in Table S2 |
| $n_{sat,ax}$ | Saturated concentration of DCVs in the axon | vesicles/µm | 0.34 |

**2.4. Age distribution of DCVs in boutons and mean age of DCVs in boutons**

In order to investigate the DCV age distribution in boutons, we followed refs. [28,29]. The compartmental system is displayed in Fig. 2. Governing equations (1)-(3) were recast as:

$$\frac{d}{dt}\mathbf{n}(t) = \mathrm{B}(\mathbf{n}(t),t)\mathbf{n}(t) + \mathbf{u}(t). \tag{21}$$

Matrix B for the case displayed in Fig. 2 is a diagonal matrix with the same elements on the main diagonal:

$$b_{ii} = -\frac{\ln(2)}{T_{1/2}} \quad (i=26,25,…,1). \tag{22}$$

The last element of vector **u** is

$$u_{26} = \left\{\min\left[h_{26}^a(n_{sat0,26} - n_{26}), j_{ax \to 26}\right] + \min\left[h_{26}^r(n_{sat0,26} - n_{26}), j_{25 \to 26}\right]\right\} / L_{26}. \tag{23}$$

The other elements of vector **u** are

$$u_i = \left\{ \min\left[ h_i^a \left( n_{sat0,i} - n_i \right), j_{i+1 \to i} \right] + \min\left[ h_i^r \left( n_{sat0,i} - n_i \right), j_{i-1 \to i} \right] \right\} / L_i \qquad (i=25,24,\ldots,2) \quad (24)$$

and

$$u_1 = \left\{ \min\left[ h_1 \left( n_{sat0,1} - n_1 \right), j_{2 \to 1} \right] \right\} / L_1. \qquad (25)$$

The fact that matrix B is diagonal means that the analyzed compartmental system does not simulate direct transfer of DCVs between the boutons. Indeed, the model displayed in Fig. 2 treats the transiting pool as a reservoir of DCVs all of which have the age of zero. For $\delta = 0$ this assumption is valid as there is no transfer of DCVs between different boutons. All the DCVs that are captured into a resident state in a bouton are eventually destroyed in that state. There is also a sufficient supply of transiting DCVs from the axon (see the analysis in section S2.6 in the Supplementary Material), such that DCV capture into the resident state is controlled by bouton capture kinetics rather than by DCV supply (see Eqs. (S1) and (S2)). However, for $\delta > 0$, a captured DCV can re-enter the transiting state and be subsequently re-captured into the resident state in one of the boutons located downstream (Fig. 2). For $\delta > 0$, treatment of the transiting pool as a reservoir of DCVs with zero age is not valid.

In order to exactly account for the age accumulated by a DCV that re-entered the transiting pool after residing in a bouton for some time, a two-concentration model, which would simulate DCV concentrations not only in the resident but also in the transiting states, needs to be developed. The difficulty of developing such a model lies in the fact that it would require the introduction of a large number of additional coefficients, which would describe transitions between transiting and resident states. In the current formulation of the model, the fluxes between the boutons contain both DCVs that have not yet been captured (their age is zero) and DCVs that have already spent some time residing in one of the boutons. Attributing previously captured DCVs to vector **u** (see Eqs. (23)-(25)) results in resetting their ages to zero. Therefore, for $\delta > 0$ the presented analysis of age distribution and mean age of DCVs is an approximation.

It should also be noted that we neglected the time it takes DCVs to travel between the boutons. A portion of DCVs that re-entered the axon ( $j_{26 \to ax}$, see Eq. (10)) can turn again, by changing the retrograde to anterograde motors, and re-enter the terminal. This situation needs to be modeled because if returning DCVs previously resided in boutons, their re-entry will affect the DCV age distribution. In type II terminals the effect of DCV return to the terminal is expected to be minor [3].

The state transition matrix, $\Phi$, was determined by solving the following matrix equation [28]:

$$\frac{d}{dt}\Phi(t,t_0) = B(\mathbf{n}(t),t)\Phi(t,t_0) \tag{26}$$

with the initial condition

$$\Phi(t_0,t_0) = I, \tag{27}$$

where I is an identity matrix. Note that $\Phi$ depends on two time variables.

Initially, the terminal did not contain any DCVs (see Eq. (11)); hence, the age density of DCVs at $t = 0$ was zero. To proceed, we need an assumption concerning the age of DCVs entering the terminal. If the DCVs had just been synthesized in the soma, they could be treated as new; however, if they had already recirculated in the terminal many times, they could be old. Ref. [3] found no DCV circulation in large axonal arbors so we assumed that all DCVs entering the terminal were new, therefore their age was set to zero. This assumption neglects the transiting time of DCVs from the soma to the terminal; therefore, the calculated DCV age should be interpreted as the age of DCVs after they enter the terminal.

The density of DCVs that entered the terminal after $t = 0$ is then calculated as:

$$\mathbf{p}(a,t) = 1_{[0,t-t_0)}(a)\Phi(t,t-a)\mathbf{u}(t-a), \tag{28}$$

where $1_{[0,t-t_0)}$ is the indicator function that is equal to 1 if $0 \le a < t - t_0$; otherwise, $1_{[0,t-t_0)}$ is equal to 0.

The mean age of DCVs in boutons, which changes as time progresses, was calculated as [30]:

$$\bar{a}_i(t) = \frac{\int_0^\infty a p_i(a,t) da}{\int_0^\infty p_i(a,t) da} \quad (i=1,\ldots,26). \tag{29}$$

$\bar{a}_i(t)$ ($i=1,\ldots,26$) were obtained by solving the following mean age system [28,30]:

$$\frac{d}{dt}\bar{\mathbf{a}}(t) = G(\mathbf{n}(t),t), \tag{30}$$

with the initial condition

$$\bar{\mathbf{a}}(0) = 0, \tag{31}$$

where $\bar{\mathbf{a}} = (\bar{a}_1, ..., \bar{a}_{26})$.

In our case G is the diagonal matrix which is defined as follows:

$$g_{ii}(t) = 1 - \frac{\bar{a}_i(t) u_i}{n_i(t)} \qquad (i=1,\ldots,26). \tag{32}$$

**2.5. Numerical solution**

We used Matlab's ODE45 solver (Matlab R2017b, MathWorks, Natick, MA, USA) to solve Eqs. (1)-(11) numerically. We set the error tolerance parameters, RelTol and AbsTol, to $10^{-6}$ and $10^{-8}$, respectively. We checked that the solution was not affected by a further decrease of RelTol and AbsTol; see Fig. S1 that shows a comparison of $n_{26}(t)$, $n_{13}(t)$, and $n_1(t)$ computed with a standard accuracy, RelTol=$10^{-6}$ and AbsTol=$10^{-8}$, with those computed with an increased accuracy, RelTol=$10^{-8}$ and AbsTol=$10^{-10}$. The computational results with standard and increased accuracy are virtually identical.

The position of a turn in the curve was determined by finding a location of large curvature. The latter was estimated by the curvature of a circle drawn through three adjacent points.

**3. Results**

Figures displaying estimated values of mass transfer coefficients characterizing the rates of DCV capture in boutons (Fig. S2a), and saturated concentrations of DCVs assuming that DCVs have infinite half-life (or infinite half-resident time) in boutons (Fig. S2b), are given in the Supplementary Material.

**3.1. Comparison of assumed steady-state concentrations with numerical results**

We checked the solution by computing steady-state concentrations in the resident state in boutons, $n_{sat,i}$, and compared them with the values assumed in Eq. (12) in the process of estimating the model parameters (Fig. 3a). It should be noted that values of $n_{sat,i}$ are not explicitly involved in the governing equations (1)-(10). A similar comparison is shown for the steady-state concentration of transiting DCVs in the axon, $n_{sat,ax}$ (Fig. 3a). By plotting concentrations of resident DCVs in boutons at various times ($t$ = 0.2, 1, and 5 h), we established that the most proximal boutons are filled first and the distal boutons are filled later in the process (Fig. 3b).

It should be noted that ref. [3] did not show the monotonic tapering-off of bouton vesicle content with distance that is output by the model. Rather, the drop-off in organelle content appeared suddenly at the furthest ends of the arbor. The model should be amended in the future to improve the agreement with the experiments in this regard.

**3.2. The case when the DCV flux from the axon to the most proximal bouton remains constant**

**3.2.1. DCV concentrations in boutons**

Fig. 4 shows how boutons reach their maximum capacity in terms of DCV accumulation. If the DCV flux from the axon to the most proximal bouton is kept constant, it takes about 8 hours for the DCV concentration in boutons to reach steady-state (Fig. 4, Table S3). An exception to this is bouton 1, where it takes about 16 hours to reach steady-state (Fig. 4b, Table S3). It takes longer to fill bouton 1 because in all other boutons, DCV capture occurs twice, first when the DCVs pass the bouton moving anterogradely and secondly when moving retrogradely (Fig. 1b). However, DCVs pass bouton 1 only once, which explains why fewer DCVs are captured and why it takes longer to fill bouton 1.

**3.2.2. DCV fluxes. The case when DCVs captured into the resident state in boutons escape and re-enter the transiting pool ($\delta = 1$)**

Anterograde (Fig. 5a) and retrograde (Fig. 5b) fluxes between the axon and the most proximal bouton and between various boutons, at the initial state ($t$=0) and at steady-state ($t \to \infty$), show that initially the anterograde DCV flux decays from the most proximal bouton to the most distal bouton (Fig. 5a). This is because fewer and fewer DCVs are left in the transiting pool as some of them are captured by boutons while traveling anterogradely. At the initial state, all retrograde fluxes are equal to zero (Fig. 4b) because it takes 300 s to change anterograde to retrograde motors in the most distal bouton (#1), hence, retrograde fluxes cannot begin before 300 s.

At steady-state, all anterograde and retrograde fluxes are equal to the flux of DCVs from the axon to the most proximal bouton, 2 DCVs/min (Fig. 5), because at $\delta = 1$ all captured DCVs reenter the transiting pool, after spending some time in the resident state. At steady-state, the rate of DCV capture is equal to the rate of DCV return to the transiting state; hence, the DCV flux is not decreasing from bouton to bouton.

It takes the anterograde and retrograde fluxes about 8 hours to reach the same steady-state value, $j_{ax \to 26}$ (Figs. 6 and 7, Table S4). It should be noted that retrograde fluxes, especially in proximal boutons, do not begin immediately, but with some delay (up to about 1 hour) which is required to fill the boutons, so that the retrograde component of the flux does not get completely depleted of DCVs before it reaches these boutons.

### 3.2.3. Distribution of DCV age in the terminal

At steady-state, the age density changes from 0 to approximately 0.4 vesicles/(μm h). The new DCVs prevail in boutons as older DCVs leave the resident state because they are destroyed in boutons (for $\delta = 0$) (Fig. 8a). Our model also makes it possible to calculate the mean age of DCVs in various boutons. At steady-state the mean DCV age in all boutons is approximately 8.66 hours (Fig. 8b, Table S5).

To investigate possible simplifications of our model, we utilized the fact that anterograde fluxes remain positive at all times (Fig. 6), while the retrograde fluxes become positive in less than one hour after the process of filling the terminal begins (Fig. 7). This fact enabled us to linearize matrix B in Eq. (21). The analysis is presented in section S2.5 of the Supplementary Material. The numerical results for the linearized case, displayed in Fig. S25, are practically identical to the results presented in Fig. 8.

Apparently, the case of $\delta = 0$ does not correctly simulate experimental findings reported in Tao et al. [3], who investigated the age of DCVs by marking the DCVs with a photoconvertible construct. Their construct switches from green to red fluorescence over a period of hours. The results reported in ref. [3] indicate that DCVs residing in distal boutons are older than those residing in proximal boutons. The fact that our model does not capture this observation (Table S5) suggests that DCVs are not destroyed in the resident state in boutons, but rather, after spending some time in the resident state, are returned to the transiting pool. As older DCVs are returned to the transiting state, for $\delta = 1$ the average age of DCVs is expected to increase from proximal to distal boutons.

**3.3. DCV fluxes. The case when DCVs captured into the resident state are destroyed in boutons ($\delta = 0$)**

Anterograde fluxes decay toward more distal boutons. This applies to the fluxes at the initial state and at steady-state, but at steady-state the decay is slower (Fig. S3a). This is because at steady-state, the boutons are filled to saturation, and DCV capture is only needed to replace the DCVs destroyed in the resident state (the rate of their destruction is controlled by the half-life of DCVs in the resident state, $T_{1/2}$). Retrograde fluxes decay from more distal to more proximal boutons (Fig. S3b). This is because as DCVs move retrogradely from more distal to more proximal DCVs, their capture continues, even at steady-state.

It takes again about 8 hours for the anterograde (Fig. S4) and retrograde (Fig. S5) fluxes to reach steady-state, as for the case with no DCV destruction (Table S4). However, fluxes in the next bouton reach a smaller value than in the previous bouton, because even at steady-state, there is now some DCV destruction in boutons which must be compensated by DCV capture from the transiting pool.

**3.4. Implications of the obtained results on DCV circulation in the terminal**

It should be noted that, unlike ref. [3], our model predicts DCV circulation in type II terminals would develop if time allowed for DCV transport to reach steady-state. This circulation is stronger if DCVs are re-released to the transiting pool after spending some time in boutons ($\delta = 1$). In this case, at steady-state, the anterograde flux of DCVs entering the terminal equals the retrograde DCV flux leaving the terminal (Fig. 5). The circulation is slightly weaker if DCVs are destroyed in boutons ($\delta = 0$, Fig. S3). Our explanation for this disagreement with results of ref. [3] is that the terminals observed in ref. [3] were not at steady-state, but were rather accumulating DCVs for later release [23]. Our results indicate that if the DCV flux from the axon to the most proximal bouton, $j_{ax \to 26}$, remains constant (at 2 vesicles/min) during the process of filling the terminal, it takes about 8 hours for the DCV fluxes to reach steady-state. In the Supplementary Material (see Figs. S6-S24 and the discussion of these figures), by simulating the DCV concentration in the axon, we investigate the situation when $j_{ax \to 26}$ depends on time, for different values of the DCV half-life in the axon. We show that a variation of $j_{ax \to 26}$ may result in a much longer time required for DCV fluxes in the axon to reach the steady-state, which may exceed the time that the *Drosophila* third instar larval stage lasts (~48 hours); the third instar were used in experiments of ref. [3].

Noteworthy, we used slightly different criteria for vesicle circulation than refs. [3,11]. For example, ref. [11] emphasized that vesicles in type Ib terminals tend to accumulate at the distal bouton early on as part of vesicle circulation. In our definition, circulation is present when a large portion of DCVs return from the most proximal bouton to the axon, in other words, when $j_{ax \to 26} / j_{26 \to ax}$ is much larger than 0. This occurs because when a large potion DCVs return to the axon, they can switch the direction of their motion to anterograde and re-enter the circulation [11]. For example, for the case displayed in Fig. 5, $j_{ax \to 26} / j_{26 \to ax}$ at steady-state is equal to unity.

**4. Discussion and future directions**

Refs. [11,12] suggested that the circulation of transiting DCVs in the terminal forms a DCV pool that can be tapped into when DCV reserves in boutons need to be replenished. Surprisingly, experiments with type III [12] and type II [3] terminals found a lack of such DCV circulation. Our results suggest that DCV circulation in a type II terminal may develop if the terminal is filled to saturation, especially if the hypothesis that all DCVs captured in the resident state in boutons are eventually released back to the transiting pool is adopted. This hypothesis is supported by the scarcity of organelle degradation machinery in terminals. The time it takes for the DCV fluxes in the axon to reach steady-state depends on the model of DCV transport in the axon and can be long, even exceeding the duration of *Drosophila's* third instar larval stage (animals in this stage were used in the DCV transport experiments). Our estimates show that the number of DCVs in the transiting state may be about one percent of the number of DCVs in the resident state. Thus, the transiting state may not provide a sufficient reserve of DCVs for replenishing boutons after a release. Instead, the DCVs traveling in the axon may play the role of a reserve source for replenishing DCV stores in boutons after a release.

Our investigation of DCV transport in the axon, presented in the Supplementary Material, suggests that the rate at which DCVs enter the axon from the soma depends on the DCV half-life in the axon. If the DCV half-life is small, a large number of DCVs must enter the axon such that some would reach the terminal. On the other hand, if the DCVs in the axon are protected from degradation (their half-life is large), then DCV transport in the axon is accomplished with minimal losses, and the rate at which DCVs enter the axon from the soma is small.

Supporting DCV circulation, the presence of which is established at least in type Ib terminals [11], requires a large amount of energy to support synthesis of DCVs that are present in the circulation. Thus, it appears that the DCV transport system is not optimized in terms of energy efficiency (to minimize the need for DCV synthesis in the soma), but rather in terms of the ability to perform a complicated task (in this case, DCV delivery) in a robust way.

A one-compartment model of the axon probably overpredicts the drop of the DCV concentration in the axon during the initial stages of filling the terminal. A more detailed model of the axon (for example, simulating the axon as consisting of several smaller sub-compartments rather than one large compartment) could be developed. However, such model would require more mass transport coefficients to simulate DCV transport between axonal sub-compartments.

**Data accessibility.** Additional data accompanying this paper are available in the Supplementary Material.

**Authors' contributions.** IAK and AVK contributed equally to the performing of computational work and article preparation.

**Competing interests.** We have no competing interests.

**Funding statement.** AVK acknowledges funding from the National Science Foundation (award CBET-1642262).

**Acknowledgments.** AVK acknowledges with gratitude the support of the Alexander von Humboldt Foundation through the Humboldt Research Award. We are grateful to Holger Metzler for careful reading of our manuscript and for his valuable comments and suggestions concerning the implementation of the method for computing the mean age of DCVs.

**Figure captions**

Fig. 1. (a) A schematic diagram showing a neuron with an axon whose terminal consists of three branches. Each branch contains 26 boutons. We numbered the boutons following the convention adopted in ref. [11]: bouton 1 is the most distal bouton while bouton 26 is the most proximal bouton. We also show 27 compartments simulated in the model (26 compartments representing the boutons and a single compartment representing the axon), as well as sizes of these compartments. (b) A magnified portion of the terminal showing boutons 20, 19, 18, and 17, fluxes between the compartments occupied by these boutons, as well as resident and transiting DCVs in the terminal. The rates of transition of anterogradely and retrogradely moving DCVs into the resident state are also shown.

Fig. 2. Schematic diagram showing transiting DCVs in the terminal and DCVs in the resident states in boutons. This compartmental representation is used for the analysis of DCV age distribution and average DCV age in the resident state in boutons. Capture of DCVs from the transiting state, destruction of DCVs in the resident state, and re-release of DCVs from the resident to the transiting state are shown by arrows. DCVs in the transiting state are assumed to have zero age.

Fig. 3. (a) Saturated DCV concentrations in the resident state in various boutons and in the transiting state in the axon. Estimated values of these concentrations, calculated using Eq. (12) and Eq. (13), are compared with numerically obtained values of saturated concentrations (obtained at $t \to \infty$). (b) Concentrations of captured DCVs in various boutons at three times: $t =$ 0.2, 1, and 5 h for the case when $j_{ax \to 26}$ is kept constant (at 2 DCVs/min).

Fig. 4. The buildup toward steady-state: concentrations of captured DCVs in various boutons. (a) Boutons 26 through 14. (b) Boutons 13 through 1. The case when $j_{ax \to 26}$ is kept constant (at 2 DCVs/min). Results are independent of $\delta$.

Fig. 5. Fluxes between the axon and the most proximal bouton and between various boutons at the initial state and at steady-state. (a) Anterograde fluxes. (b) Retrograde fluxes. The case when $j_{ax \to 26}$ is kept constant (at 2 DCVs/min), $\delta = 1$ (which refers to the case when all captured DCVs eventually reenter the transiting pool).

Fig. 6. The buildup toward steady-state: the flux from the axon to the most proximal bouton and anterograde fluxes between various boutons. (a) Fluxes ax→26 through 1→14. (b) Fluxes 14→13

through 2→1. The case when $j_{ax \to 26}$ is kept constant (at 2 DCVs/min), $\delta = 1$ (which refers to the case when all captured DCVs eventually reenter the transiting pool).

Fig. 7. The buildup toward steady-state: retrograde fluxes between various boutons and the flux from the most proximal bouton to the axon. (a) Fluxes 1→2 through 13→14. (b) Fluxes 14→15 through 26→ax. The case when $j_{ax \to 26}$ is kept constant (at 2 DCVs/min), $\delta = 1$ (which refers to the case when all captured DCVs eventually reenter the transiting pool).

Fig. 8. (a) Age density of DCVs in various boutons at steady-state. (b) Mean age of resident DCVs in various boutons versus time. The case when $j_{ax \to 26}$ is kept constant (at 2 DCVs/min). The presented age of DCV analysis applies to the situation characterized by $\delta = 0$ (which refers to the case when all captured DCVs are eventually destroyed in boutons).

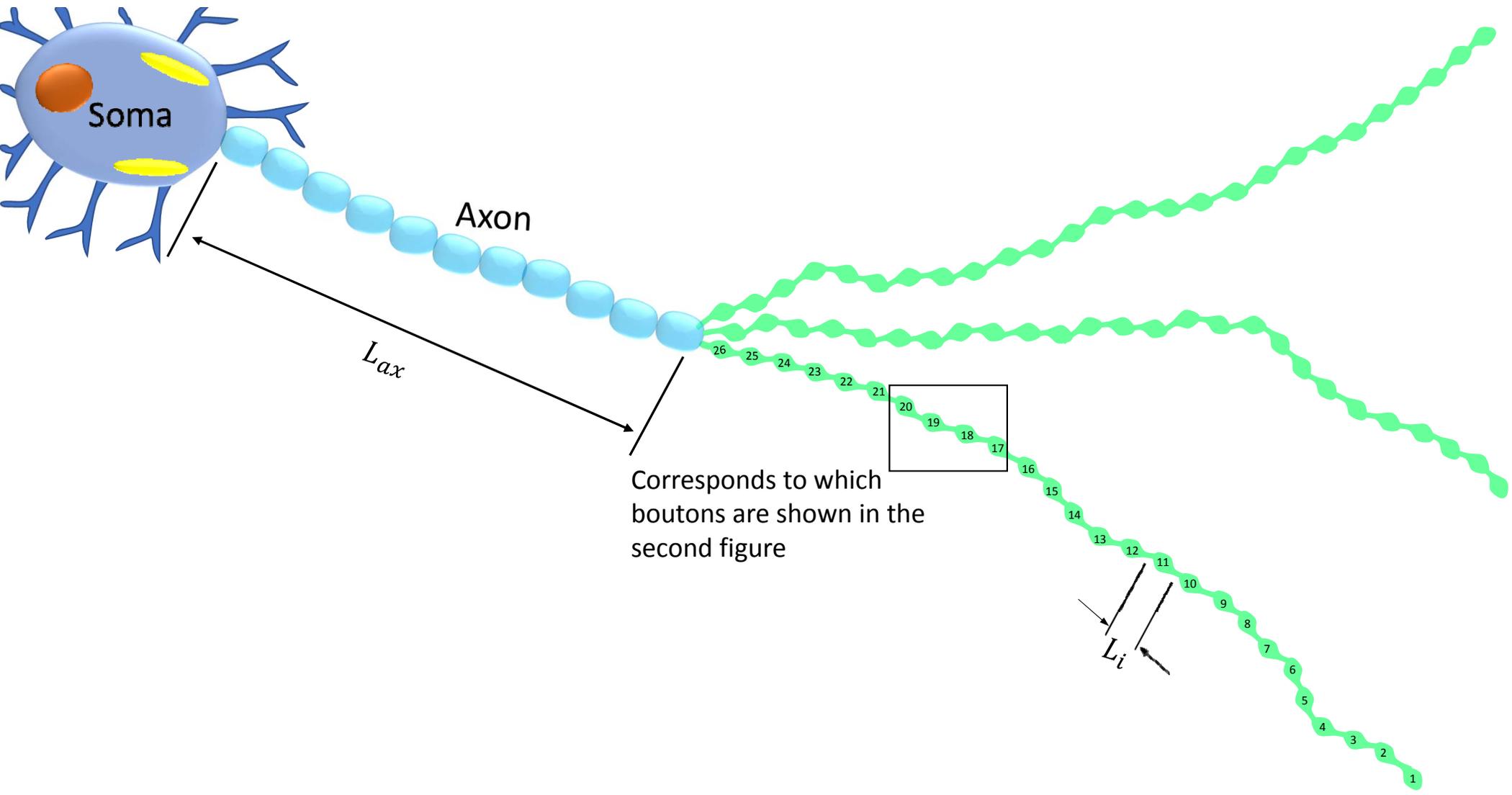

Figure 1a

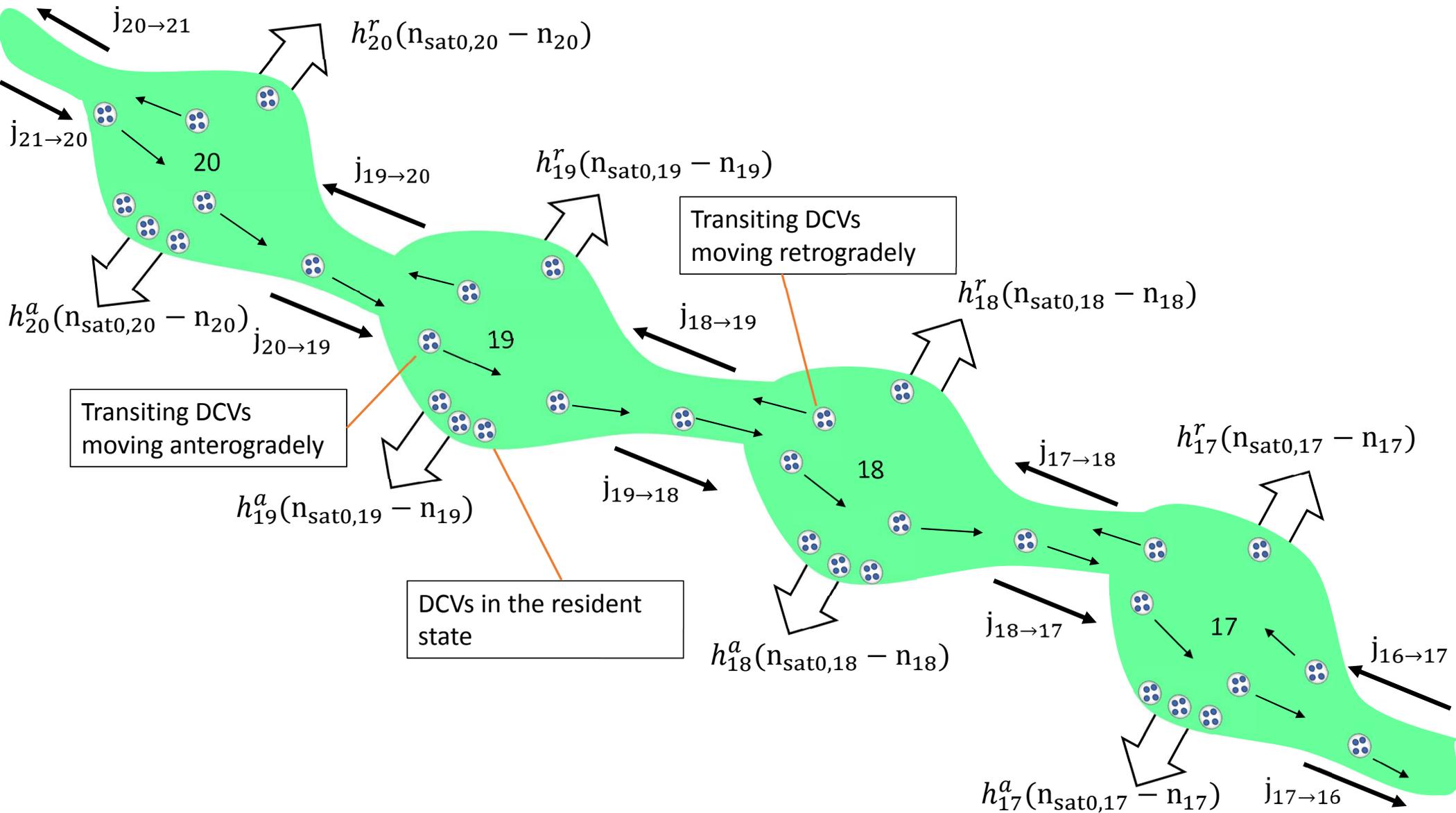

Figure 1b

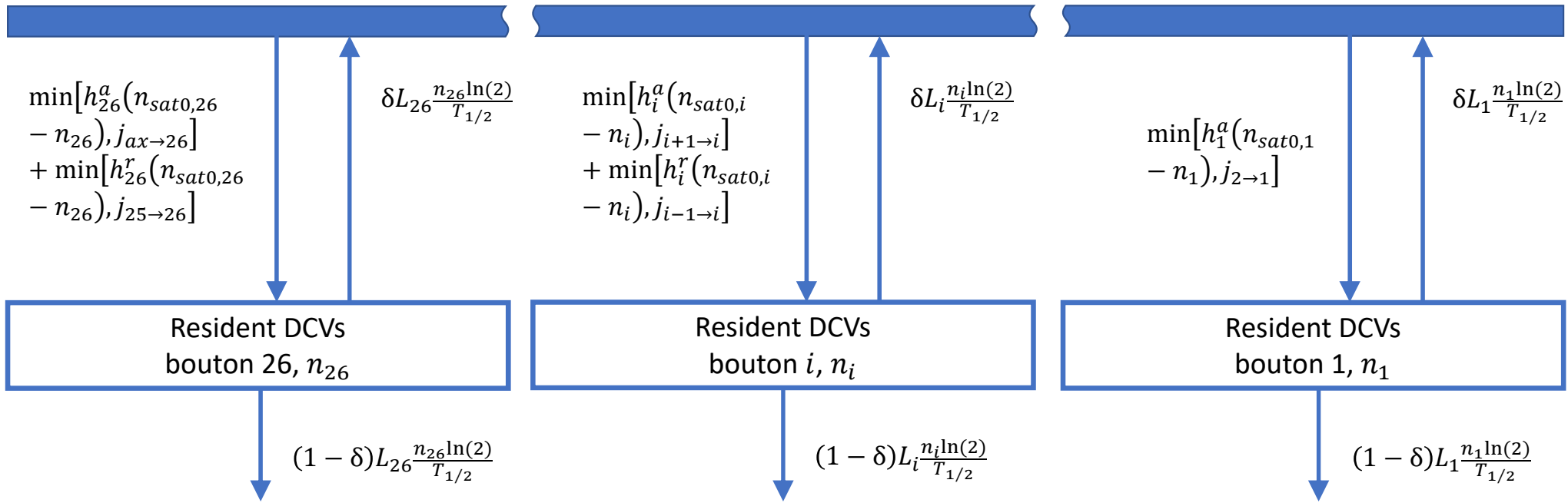

Figure 2

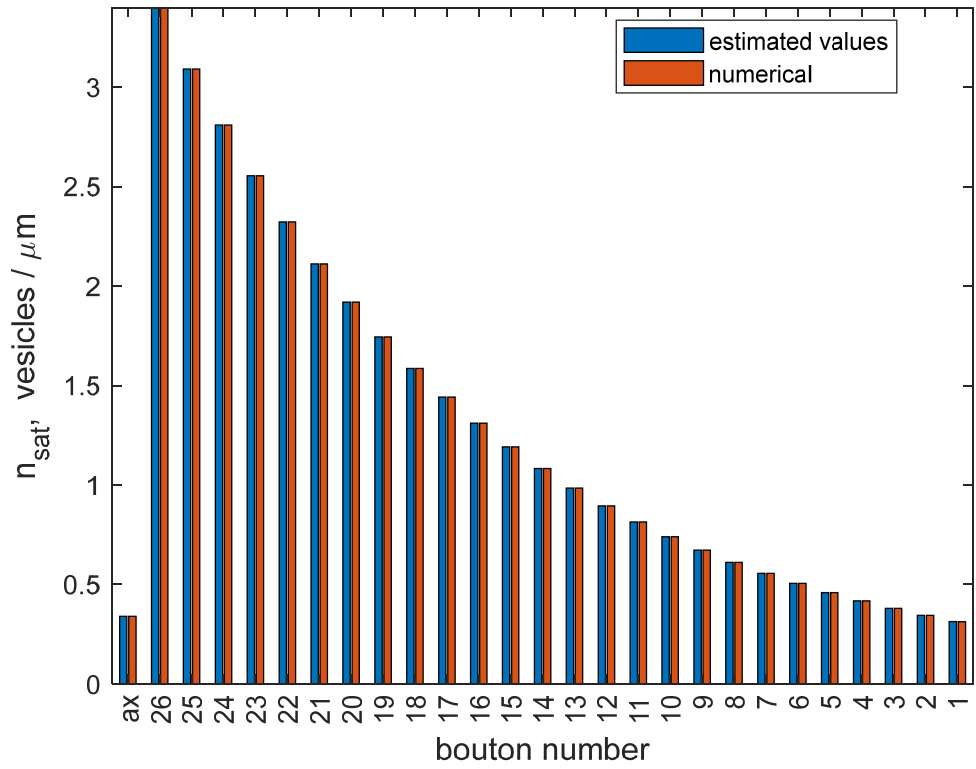

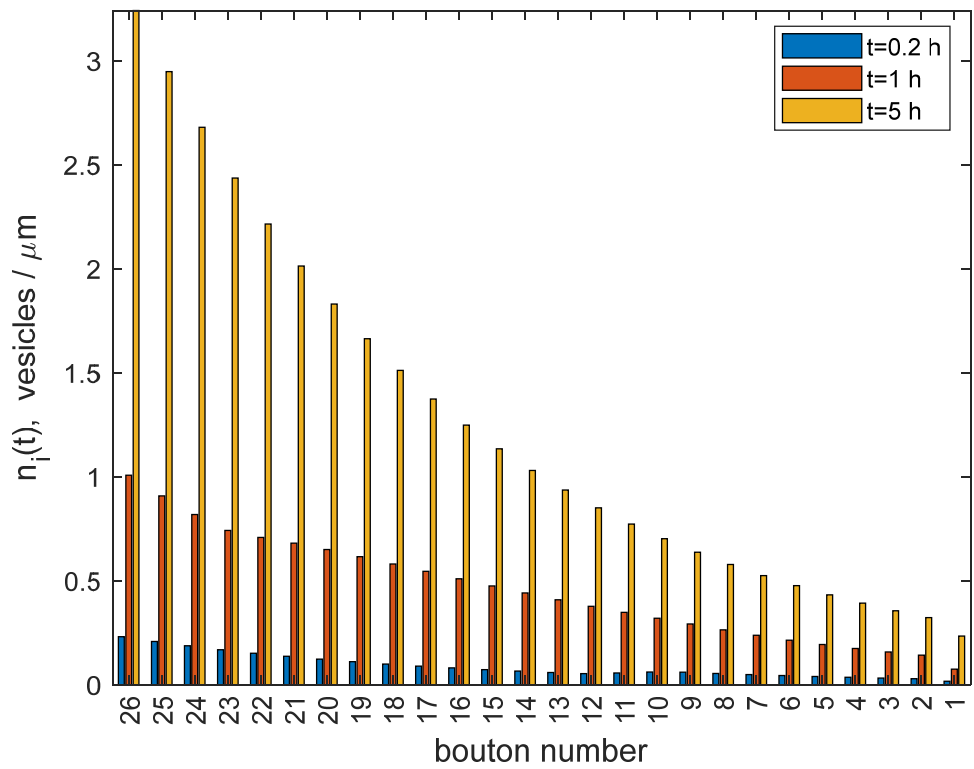

Figure 3

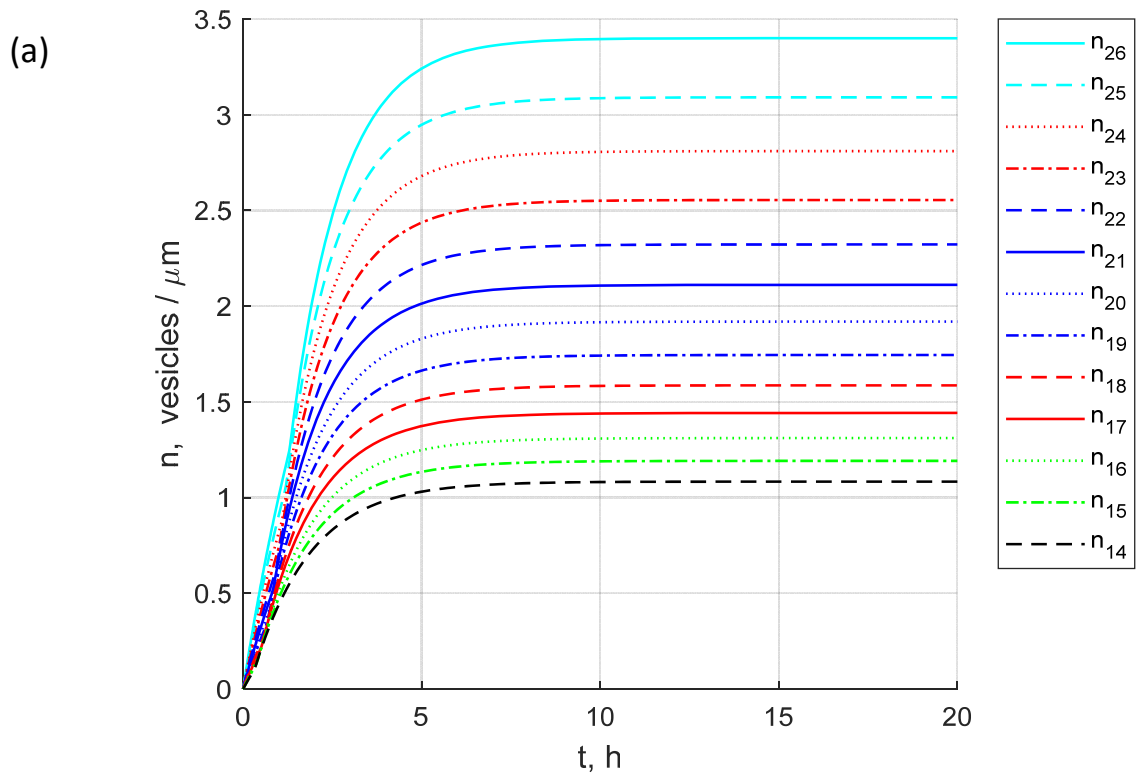

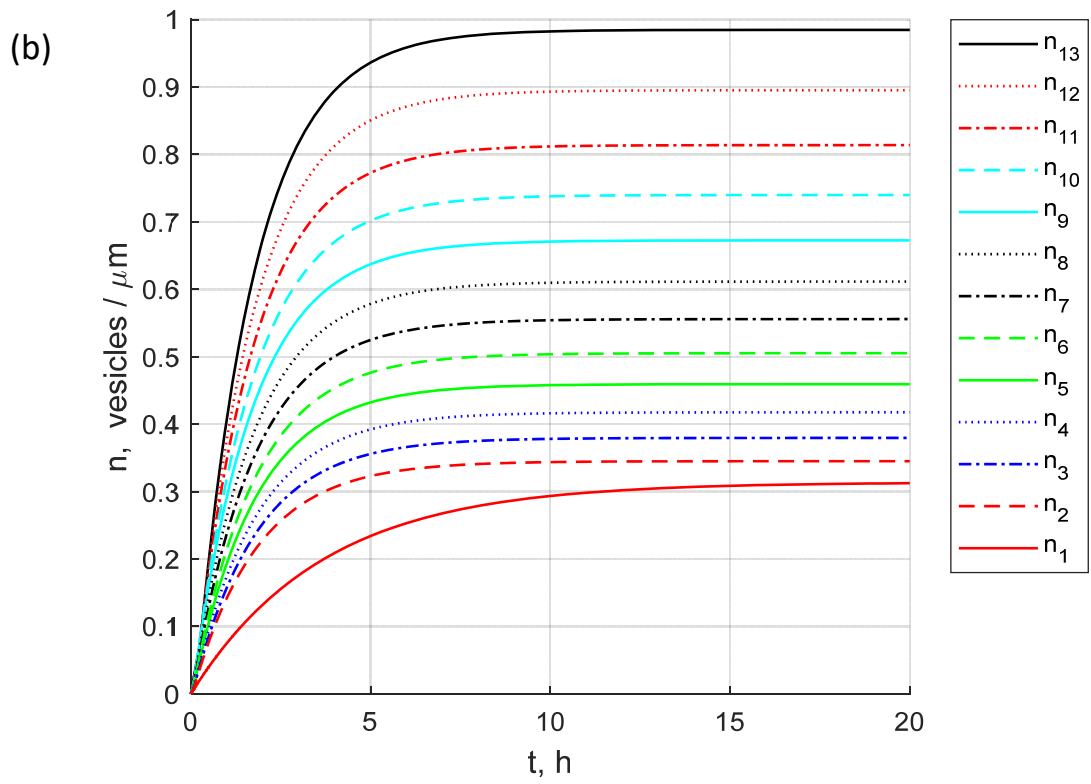

Figure 4

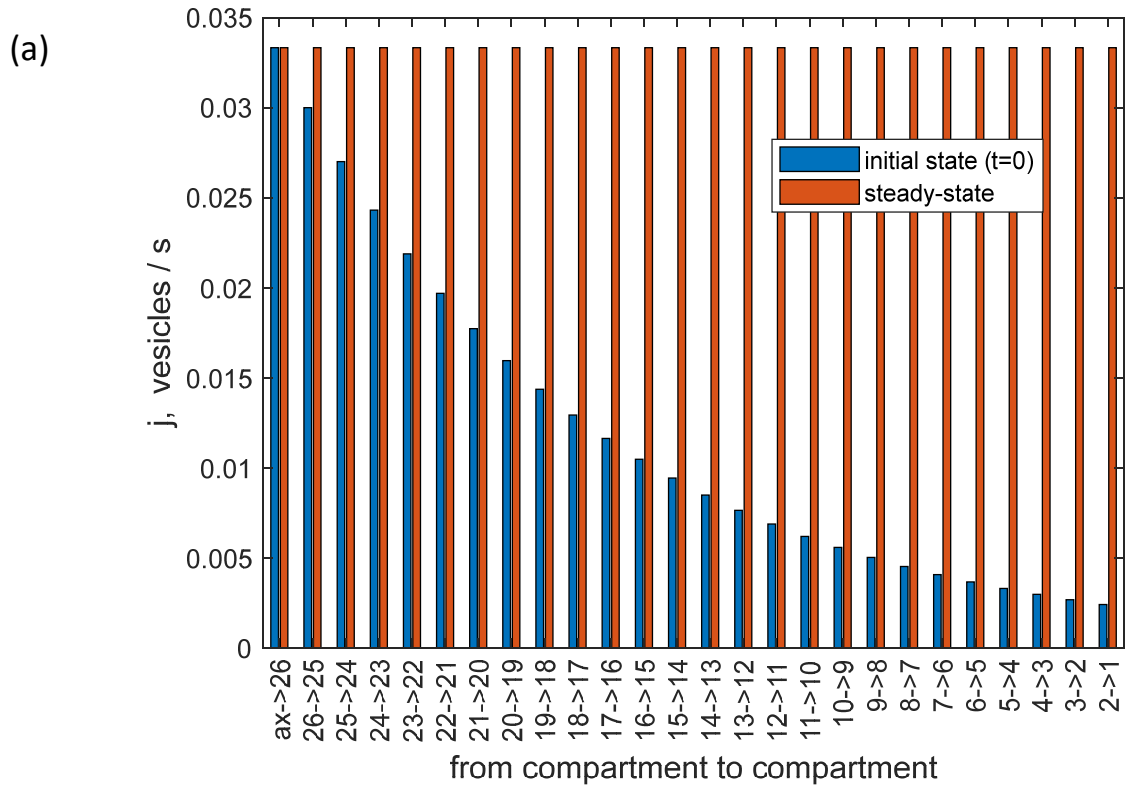

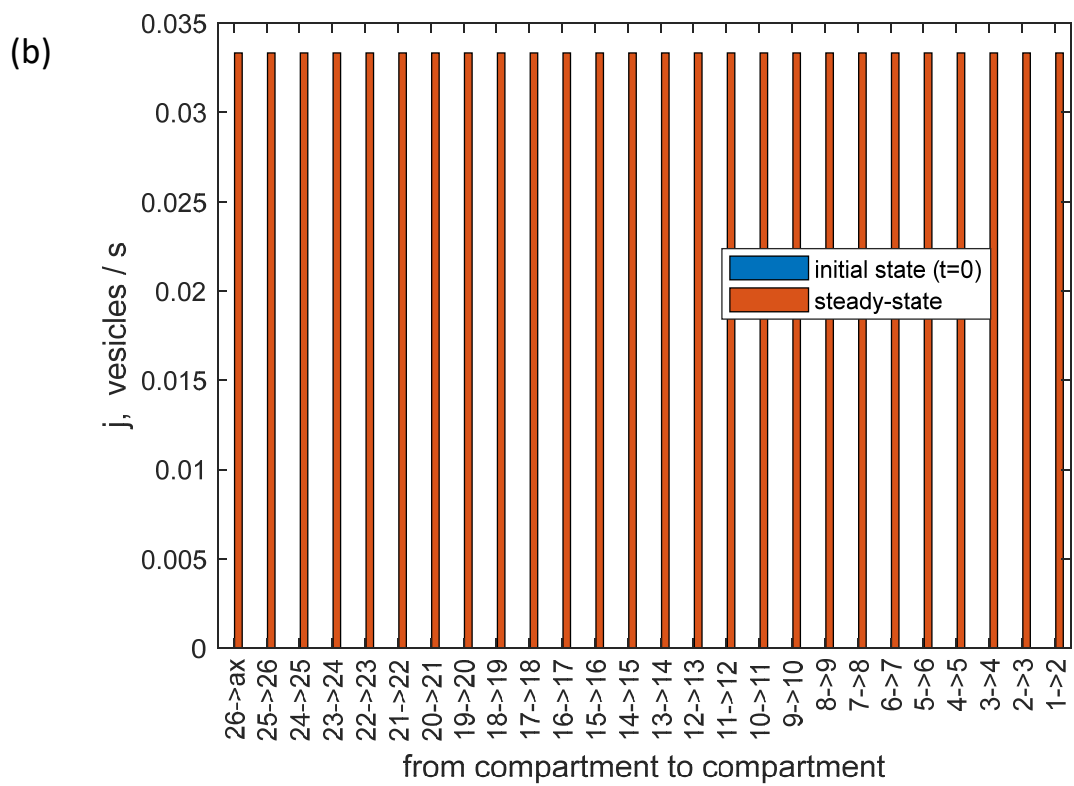

Figure 5

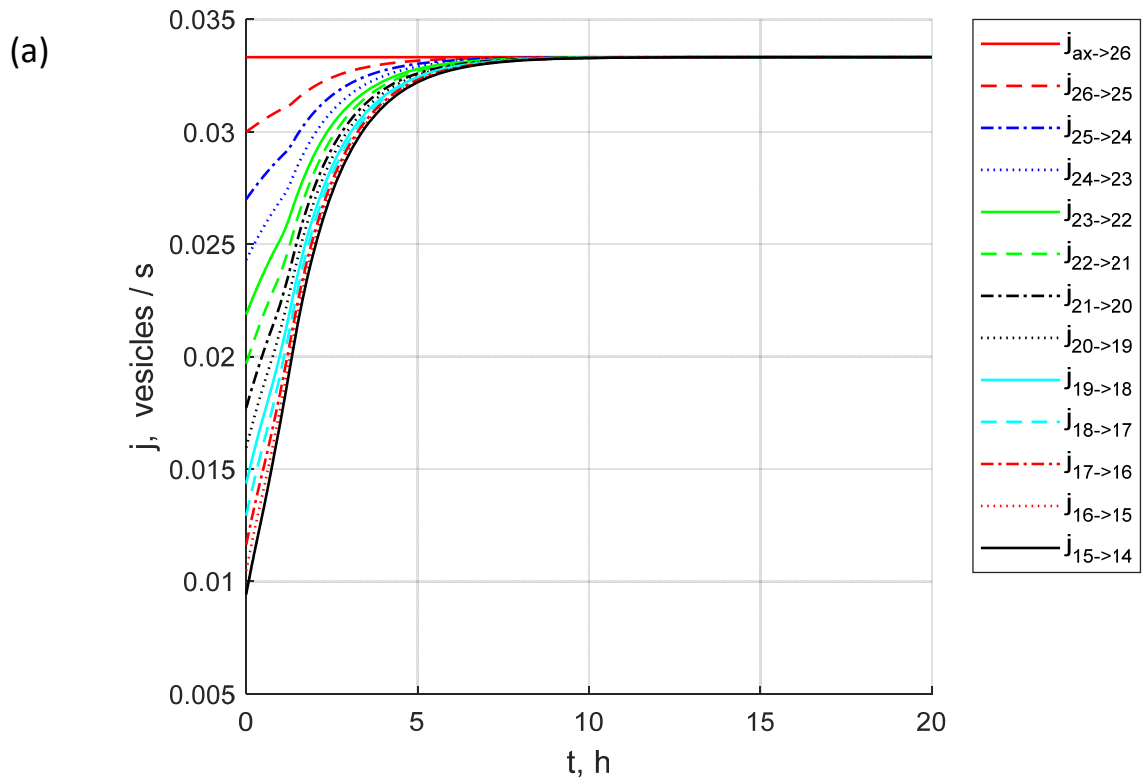

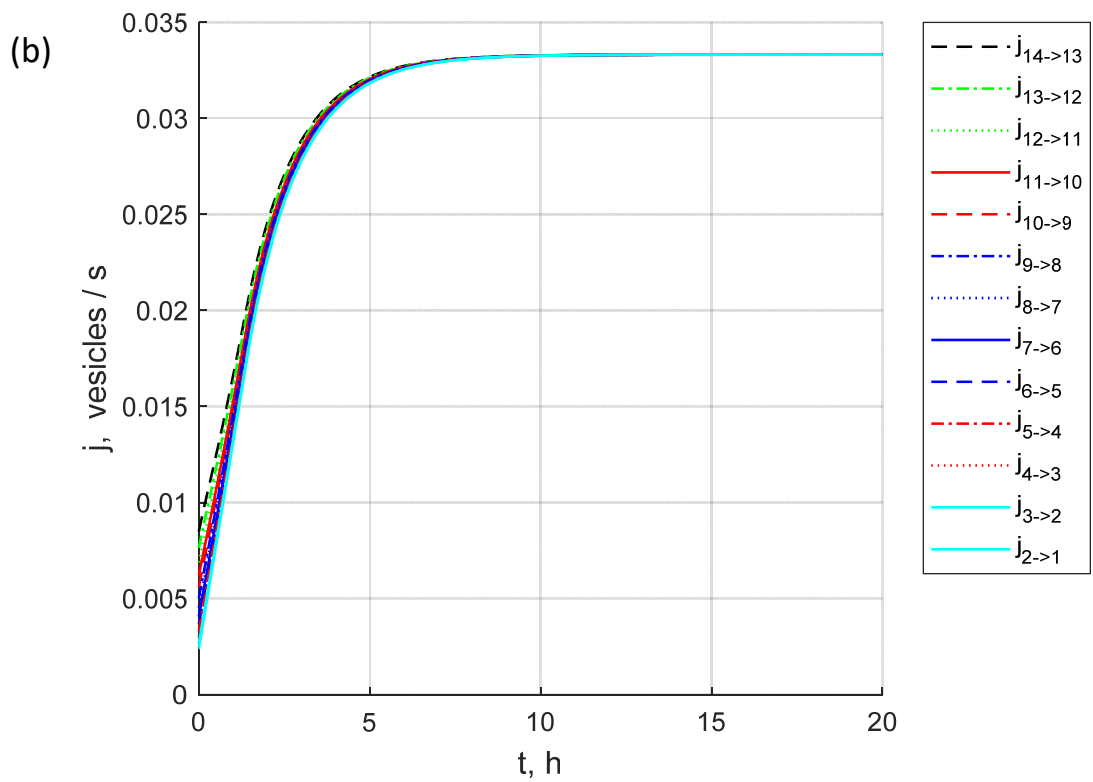

Figure 6

(a)

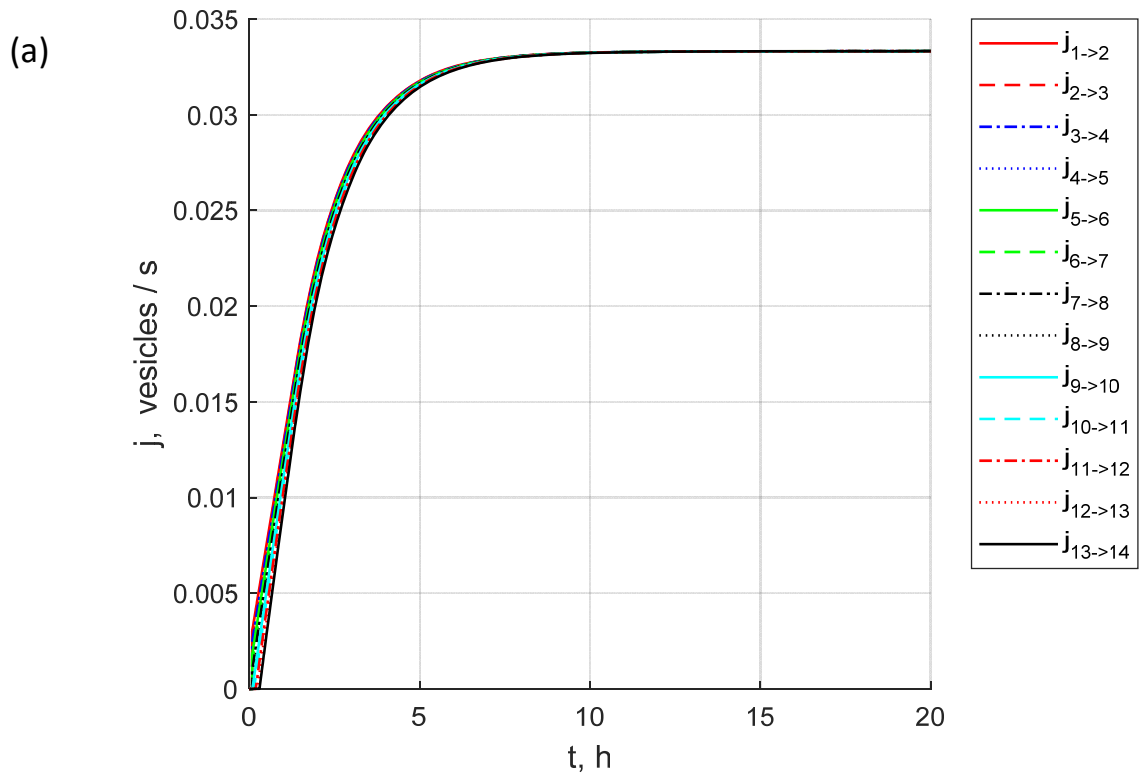

(b)

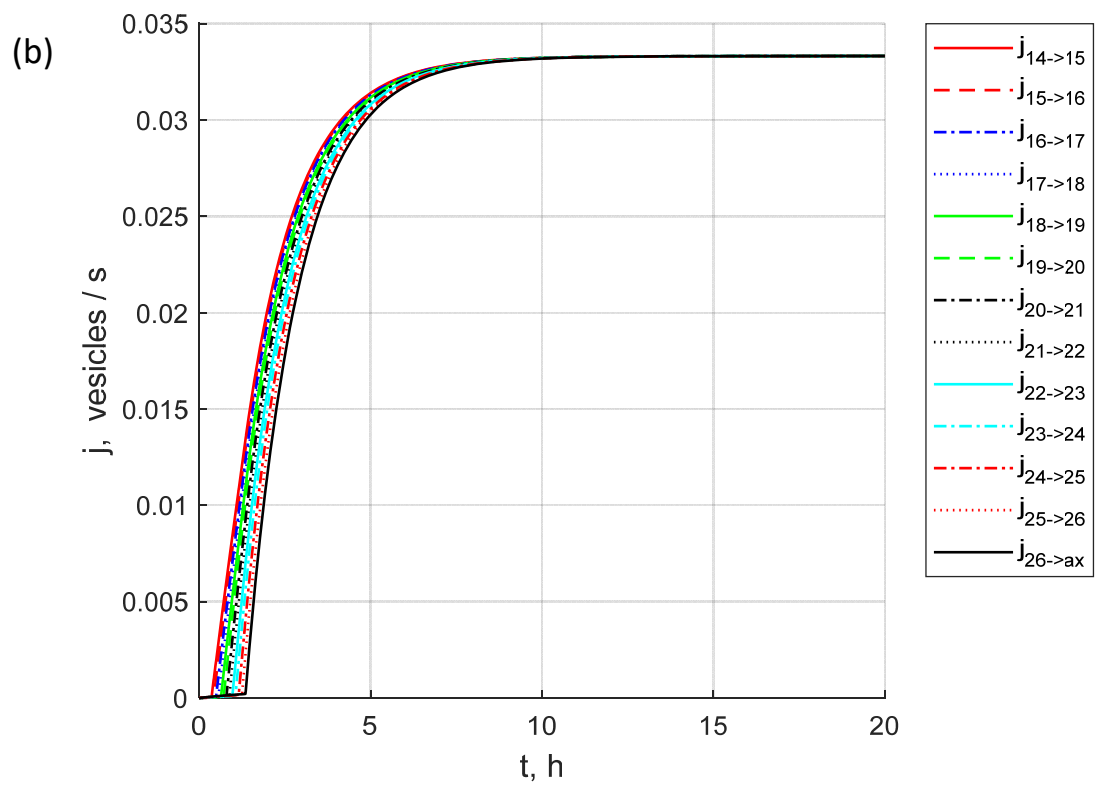

Figure 7

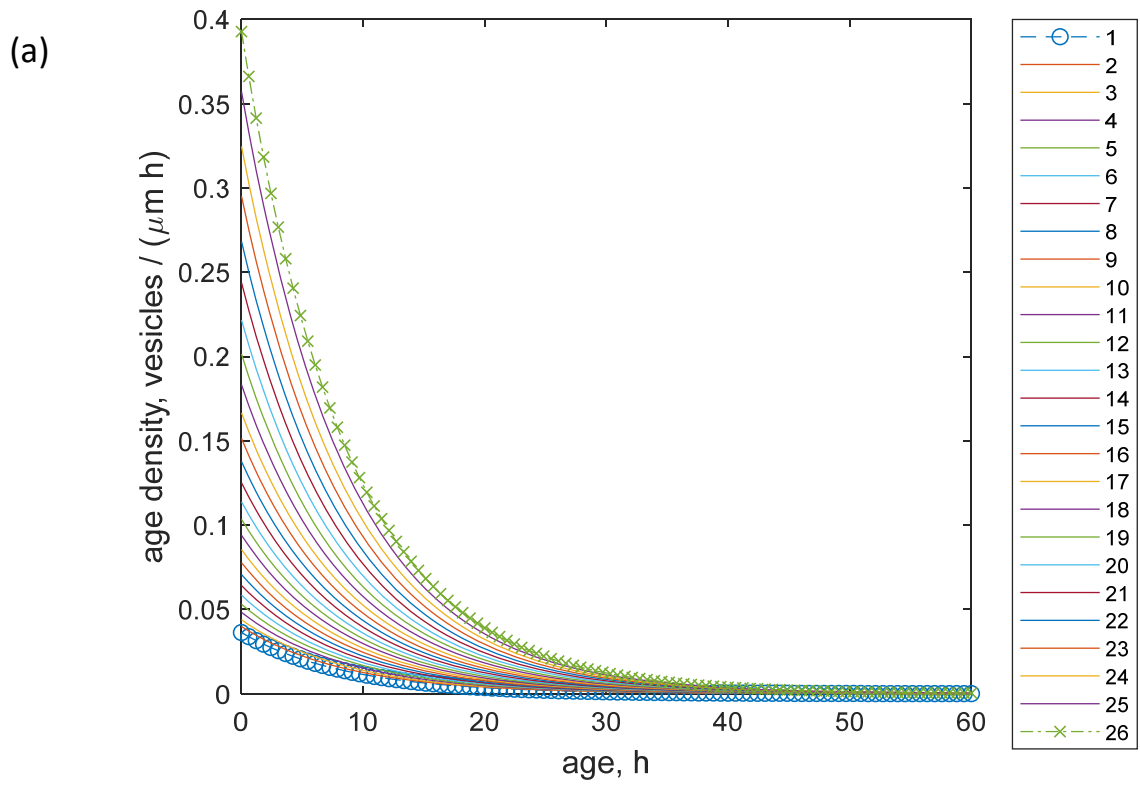

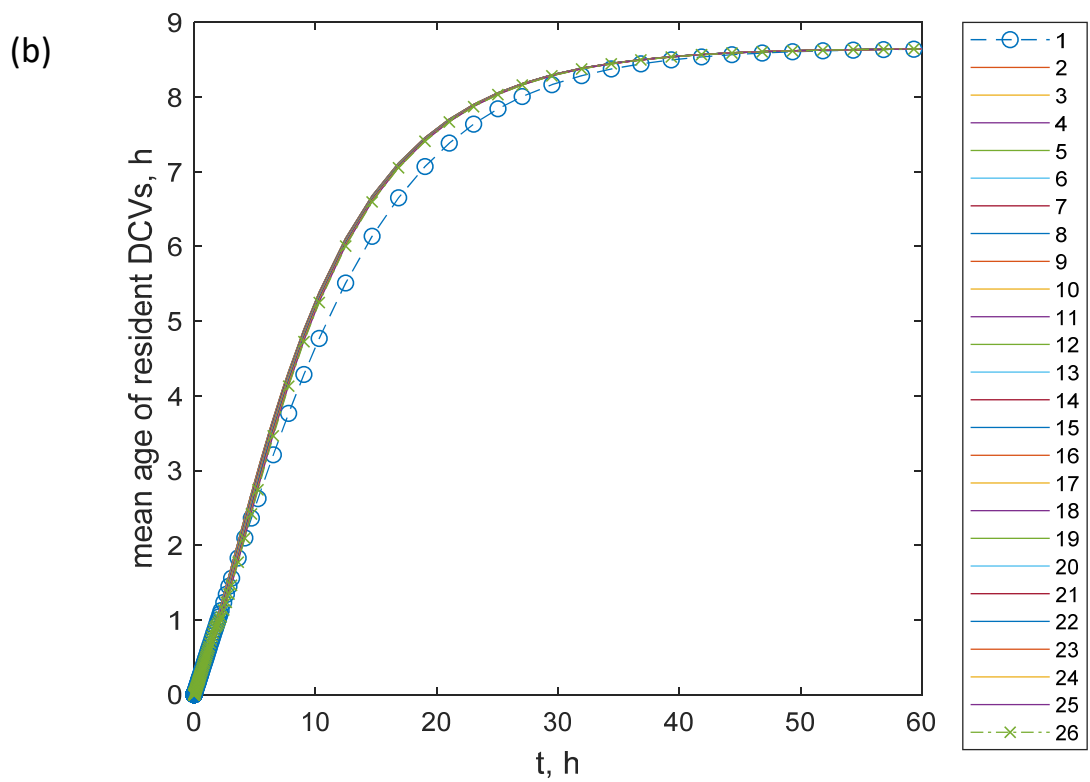

Figure 8

# Modeling transport and mean age of dense core vesicles in large axonal arbors

I. A. Kuznetsov[(a), (b)] and A. V. Kuznetsov[(c)]

[(a)]Perelman School of Medicine, University of Pennsylvania, Philadelphia, PA 19104, USA

[(b)]Department of Bioengineering, University of Pennsylvania, Philadelphia, PA 19104, USA

[(c)]Dept. of Mechanical and Aerospace Engineering, North Carolina State University, Raleigh, NC 27695-7910, USA; e-mail: avkuznet@ncsu.edu

## Supplementary Material

### S1. Supplementary tables

Table S1. Mass transfer coefficients characterizing the rates of capture of DCVs into the resident state. In estimating these coefficients, we assumed that $h_i^a = h_i^r = h_i$ for $i=2,\ldots,26$ (in bouton 1 there is no difference between anterograde and retrograde coefficients because DCVs turn around at this bouton). The units of all $h_i$ in Table S1 are μm/s.

| $h_1$ | $h_2$ | $h_3$ | $h_4$ | $h_5$ | $h_6$ | $h_7$ |
|---|---|---|---|---|---|---|
| $4.41 \times 10^{-4}$ | $6.09 \times 10^{-4}$ | $6.17 \times 10^{-4}$ | $6.25 \times 10^{-4}$ | $6.33 \times 10^{-4}$ | $6.41 \times 10^{-4}$ | $6.49 \times 10^{-4}$ |

| $h_8$ | $h_9$ | $h_{10}$ | $h_{11}$ | $h_{12}$ | $h_{13}$ | $h_{14}$ |
|---|---|---|---|---|---|---|
| $6.57 \times 10^{-4}$ | $6.65 \times 10^{-4}$ | $6.73 \times 10^{-4}$ | $6.82 \times 10^{-4}$ | $6.90 \times 10^{-4}$ | $6.99 \times 10^{-4}$ | $7.08 \times 10^{-4}$ |

| $h_{15}$ | $h_{16}$ | $h_{17}$ | $h_{18}$ | $h_{19}$ | $h_{20}$ | $h_{21}$ |
|---|---|---|---|---|---|---|
| $7.16 \times 10^{-4}$ | $7.25 \times 10^{-4}$ | $7.34 \times 10^{-4}$ | $7.43 \times 10^{-4}$ | $7.52 \times 10^{-4}$ | $7.62 \times 10^{-4}$ | $7.71 \times 10^{-4}$ |

| $h_{22}$ | $h_{23}$ | $h_{24}$ | $h_{25}$ | $h_{26}$ |
|---|---|---|---|---|
| $7.80 \times 10^{-4}$ | $7.90 \times 10^{-4}$ | $7.99 \times 10^{-4}$ | $8.09 \times 10^{-4}$ | $8.19 \times 10^{-4}$ |

Table S2. Saturated concentrations of DCVs in boutons at infinite DCV half-life or at infinite DCV residence time. The units of $n_{sat0,i}$ in Table S2 are vesicles/μm.

| $n_{sat0,1}$ | $n_{sat0,2}$ | $n_{sat0,3}$ | $n_{sat0,4}$ | $n_{sat0,5}$ | $n_{sat0,6}$ | $n_{sat0,7}$ |
|---|---|---|---|---|---|---|
| 0.542 | 0.436 | 0.478 | 0.525 | 0.576 | 0.632 | 0.693 |

| $n_{sat0,8}$ | $n_{sat0,9}$ | $n_{sat0,10}$ | $n_{sat0,11}$ | $n_{sat0,12}$ | $n_{sat0,13}$ | $n_{sat0,14}$ |
|---|---|---|---|---|---|---|
| 0.761 | 0.835 | 0.916 | 1.01 | 1.10 | 1.21 | 1.33 |

| $n_{sat0,15}$ | $n_{sat0,16}$ | $n_{sat0,17}$ | $n_{sat0,18}$ | $n_{sat0,19}$ | $n_{sat0,20}$ | $n_{sat0,21}$ |
|---|---|---|---|---|---|---|
| 1.46 | 1.60 | 1.76 | 1.93 | 2.12 | 2.32 | 2.55 |

| $n_{sat0,22}$ | $n_{sat0,23}$ | $n_{sat0,24}$ | $n_{sat0,25}$ | $n_{sat0,26}$ |
|---|---|---|---|---|
| 2.80 | 3.07 | 3.37 | 3.70 | 4.07 |

Table S3. Approximate time required to reach steady-state concentration in various boutons for various models of DCV transport in the axon. We assumed that steady-state is reached when the DCV concentration in a particular bouton reaches 99% of $n_{sat,i}$, which is defined in Eq. (12).

| Axonal transport model | $t_\infty$ for $n_{ax}$, h | $t_\infty$ for $n_{26}$, h | $t_\infty$ for $n_{13}$, h | $t_\infty$ for $n_1$, h |
|---|---|---|---|---|
| $j_{ax \to 26} = 0.0333$ vesicles/s | N/A | 7.18 | 7.58 | 16.80 |
| $j_{ax \to 26}$ is modeled by Eq. (5), $\delta = 1$, $T_{1/2,ax} = 1 \times T_{1/2}$ | 87.41 | 46.61 | 47.47 | 59.93 |
| $j_{ax \to 26}$ is modeled by Eq. (5), $\delta = 1$, $T_{1/2,ax} = 100 \times T_{1/2}$ | 8,396 | 4,419 | 3,039 | 2,725 |
| $j_{ax \to 26}$ is modeled by Eq. (5), $\delta = 0$, $T_{1/2,ax} = 1 \times T_{1/2}$ | 55.46 | 21.12 | 15.95 | 25.11 |
| $j_{ax \to 26}$ is modeled by Eq. (5), $\delta = 0$, $T_{1/2,ax} = 100 \times T_{1/2}$ | 3,697 | 52.32 | 20.09 | 28.41 |

Table S4. Approximate time required to reach steady-state for anterograde and retrograde fluxes in various boutons for various models of DCV transport in the axon. We assumed that steady-state is reached when the corresponding flux reaches 99% of its steady-state value.

| Axonal transport model | $t_\infty$ for $j_{ax \to 26}$, h | $t_\infty$ for $j_{2 \to 1}$, h | $t_\infty$ for $j_{1 \to 2}$, h | $t_\infty$ for $j_{26 \to ax}$, h |
|---|---|---|---|---|
| $j_{ax \to 26} = 0.0333$ vesicles/s, $\delta = 1$ | N/A | 7.29 | 7.44 | 8.47 |
| $j_{ax \to 26} = 0.0333$ vesicles/s, $\delta = 0$ | N/A | 7.39 | 7.49 | 8.87 |
| $j_{ax \to 26}$ is modeled by Eq. (5), $\delta = 1$, $T_{1/2,ax} = 1 \times T_{1/2}$ | 87.41 | 87.41 | 87.41 | 87.41 |
| $j_{ax \to 26}$ is modeled by Eq. (5), $\delta = 1$, $T_{1/2,ax} = 100 \times T_{1/2}$ | 8,396 | 8,396 | 8,396 | 8,396 |
| $j_{ax \to 26}$ is modeled by Eq. (5), $\delta = 0$, $T_{1/2,ax} = 1 \times T_{1/2}$ | 55.46 | 56.99 | 57.03 | 58.91 |
| $j_{ax \to 26}$ is modeled by Eq. (5), $\delta = 0$, $T_{1/2,ax} = 100 \times T_{1/2}$ | 3,697 | 3,852 | 3,855 | 4,044 |

Table S5. The mean age of resident DCVs in various boutons at steady-state for the case when $j_{ax \to 26}$ is kept constant (at 2 DCVs/min). The case of $\delta = 0$, which refers to the situation when all captured DCVs are eventually destroyed in boutons.

| Mean age in bouton 1, h | Mean age in bouton 13, h | Mean age in bouton 26, h |
|---|---|---|
| 8.66 | 8.66 | 8.66 |

Table S6. The time when switching between two different transport regimes occurs: (i) the regime when DCV capture in boutons exceeds the rate of DCV synthesis in the soma and (ii) the regime when the terminal is filled to saturation, and the DCV concentration in the axon recovers to its steady-state value. This time is shown by a rhombus in Figs. S8, S12, S17, and S21.

| Axonal transport model | $t_{switch}$, h |
|---|---|
| $j_{ax \to 26}$ is modeled by Eq. (5), $\delta = 1$, $T_{1/2,ax} = 1 \times T_{1/2}$ | 46.05 |
| $j_{ax \to 26}$ is modeled by Eq. (5), $\delta = 1$, $T_{1/2,ax} = 100 \times T_{1/2}$ | 4,423 |
| $j_{ax \to 26}$ is modeled by Eq. (5), $\delta = 0$, $T_{1/2,ax} = 1 \times T_{1/2}$ | 17.86 |
| $j_{ax \to 26}$ is modeled by Eq. (5), $\delta = 0$, $T_{1/2,ax} = 100 \times T_{1/2}$ | 56.13 |

## S2. Supplementary figures

### S2.1. Independence of the solution of the values of error tolerance parameters, RelTol and AbsTol

Independence of the solution of the values of error tolerance parameters, RelTol and AbsTol, is demonstrated in Fig. S1.

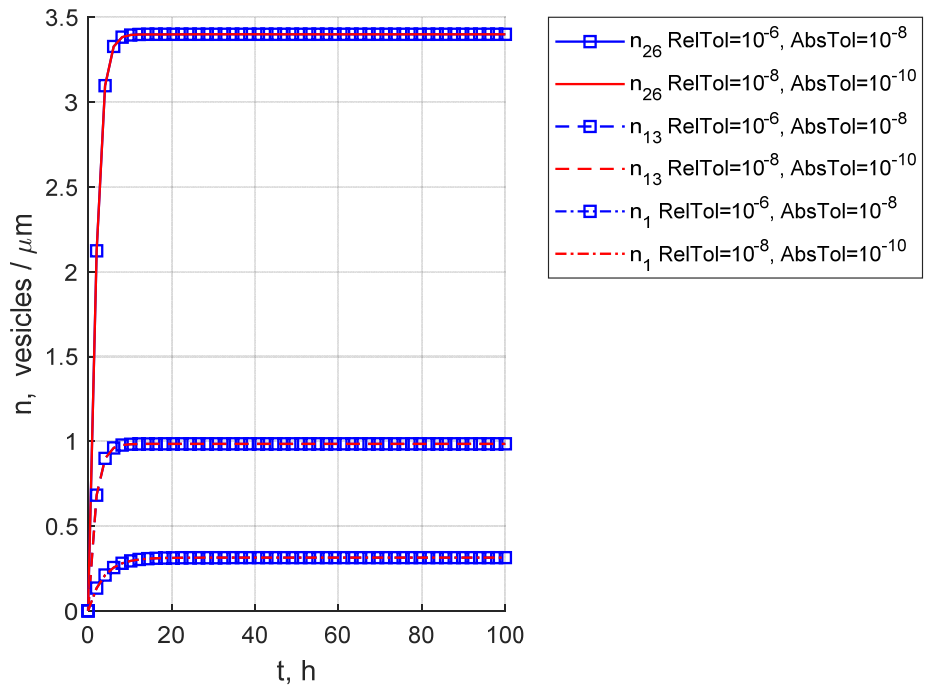

Fig. S1. A comparison of the concentration of DCVs in three boutons, 26, 13, and 1, computed with a standard accuracy and with an increased accuracy.

**S2.2. Estimated values of mass transfer coefficients and saturated DCV concentrations in boutons assuming infinite DCV residence time**

Estimated values of mass transfer coefficients characterizing the rates of DCV capture into the resident state decrease from the most proximal (#26) to the most distal bouton (#1) (Fig. S2a). The decrease is explained by lower rates of DCV capture in more proximal boutons due to lower DCV concentrations in these boutons at saturation (Eq. (12)). A much lower value of the mass transfer coefficient for the most distal bouton, $h_1$, than for other boutons, is explained by the fact that DCVs pass bouton 1 only once, and thus there is no double-capture (from anterogradely and retrogradely moving transiting pools).

The concentrations of DCVs at steady-state for infinite DCV residence time (or infinite half-life) decrease from the most proximal (#26) to the most distal bouton (#1) (Fig. S2b). This is again due to the decreased steady-state DCV concentration in proximal boutons (Eq. (12)). A slightly larger value of $n_{sat0,1}$ is to offset a significant decrease in $h_1$.

The values of parameters $h_i$ and $n_{sat0,i}$ that are displayed in Figs. S2a and S2b, respectively, are the same for all cases simulated in this paper.

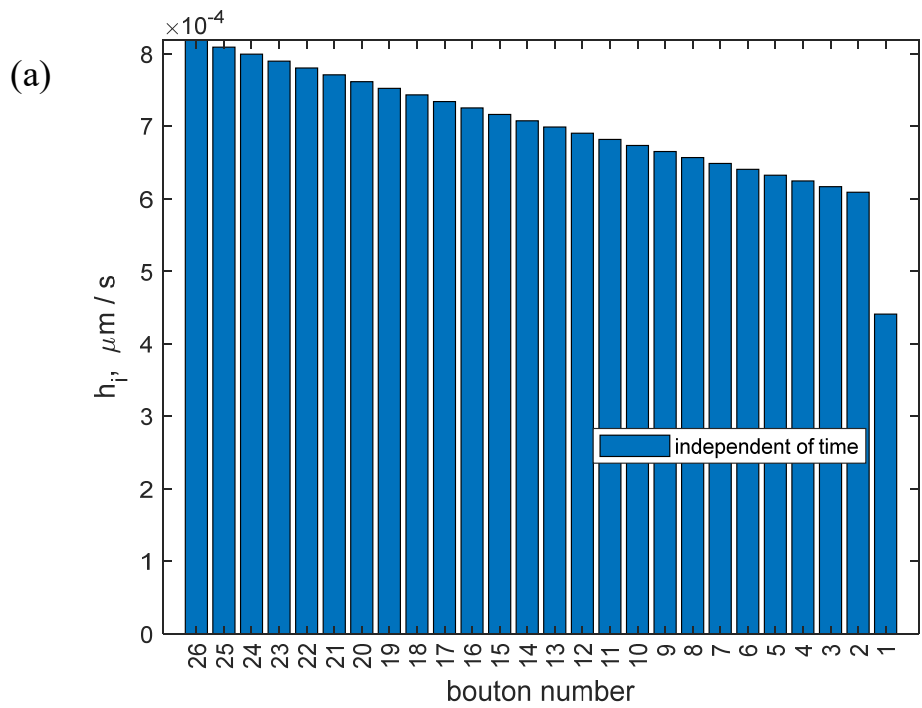

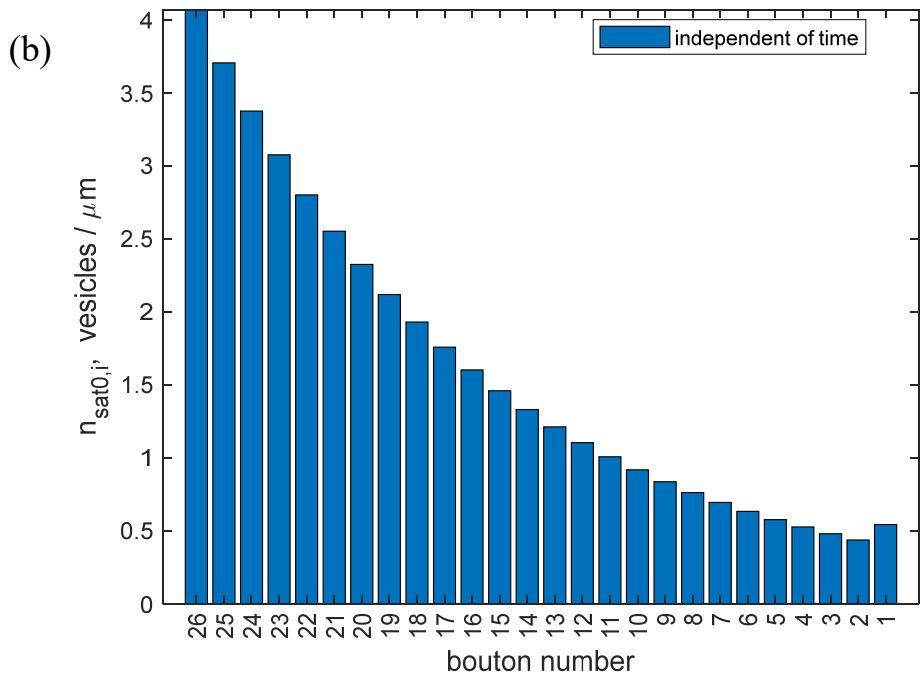

Fig. S2. Estimated values of parameters (these values are the same for all cases investigated in this paper). (a) Mass transfer coefficients characterizing the rates of DCV capture into the resident state in various boutons. (b) Saturated concentrations of resident DCVs in various boutons at infinite DCV residence time (or infinite half-life).

**S2.3. The case when the DCV flux from the axon to the most proximal bouton is constant, and DCVs captured into the resident state are destroyed in boutons ($\delta = 0$)**

In this series of simulations, we show DCV fluxes for the case of $j_{ax \to 26} = 2$ DCVs/min and $\delta = 0$.

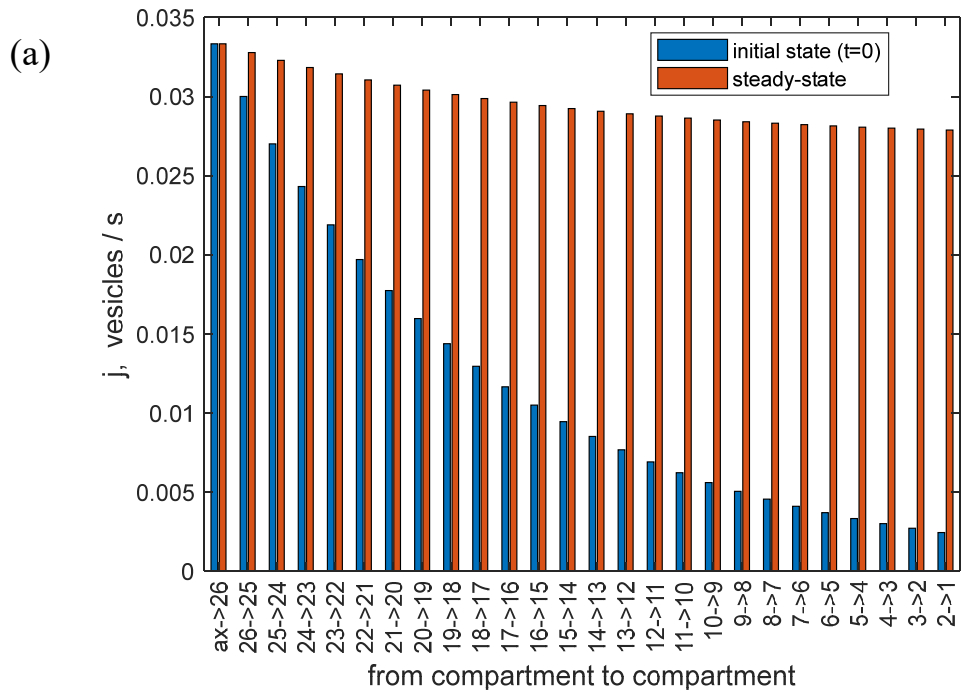

(a)

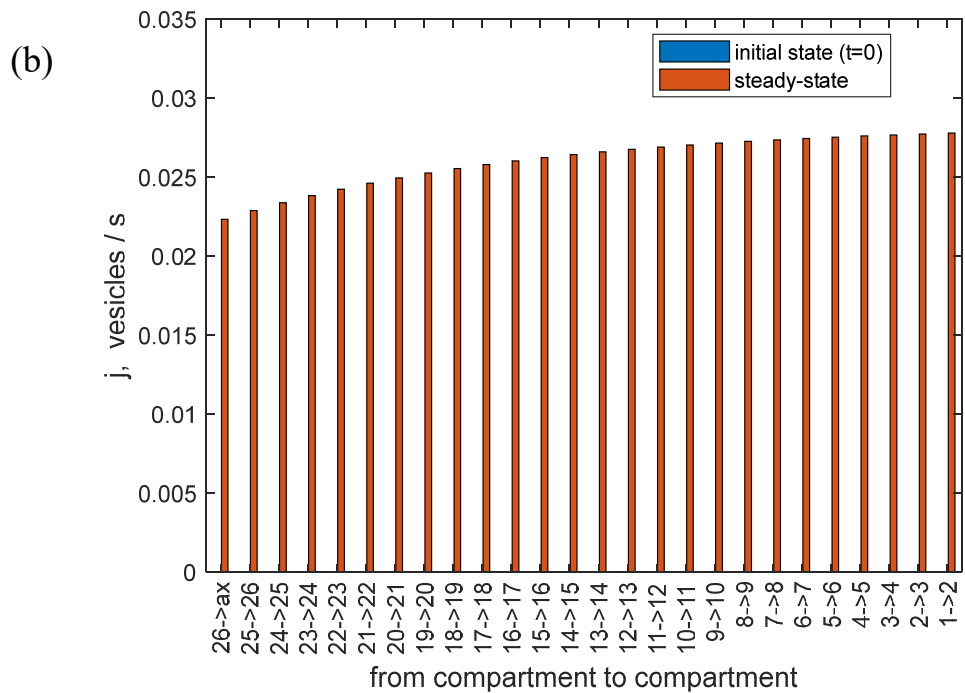

Fig. S3. Fluxes between the axon and the most proximal bouton and between various boutons at the initial state and at steady-state. (a) Anterograde fluxes. (b) Retrograde fluxes. The case when $j_{ax \to 26}$ is kept constant (at 2 DCVs/min). The case when $j_{ax \to 26}$ is kept constant (at 2 DCVs/min), $\delta = 0$ (which refers to the case when all captured DCVs are destroyed in boutons).

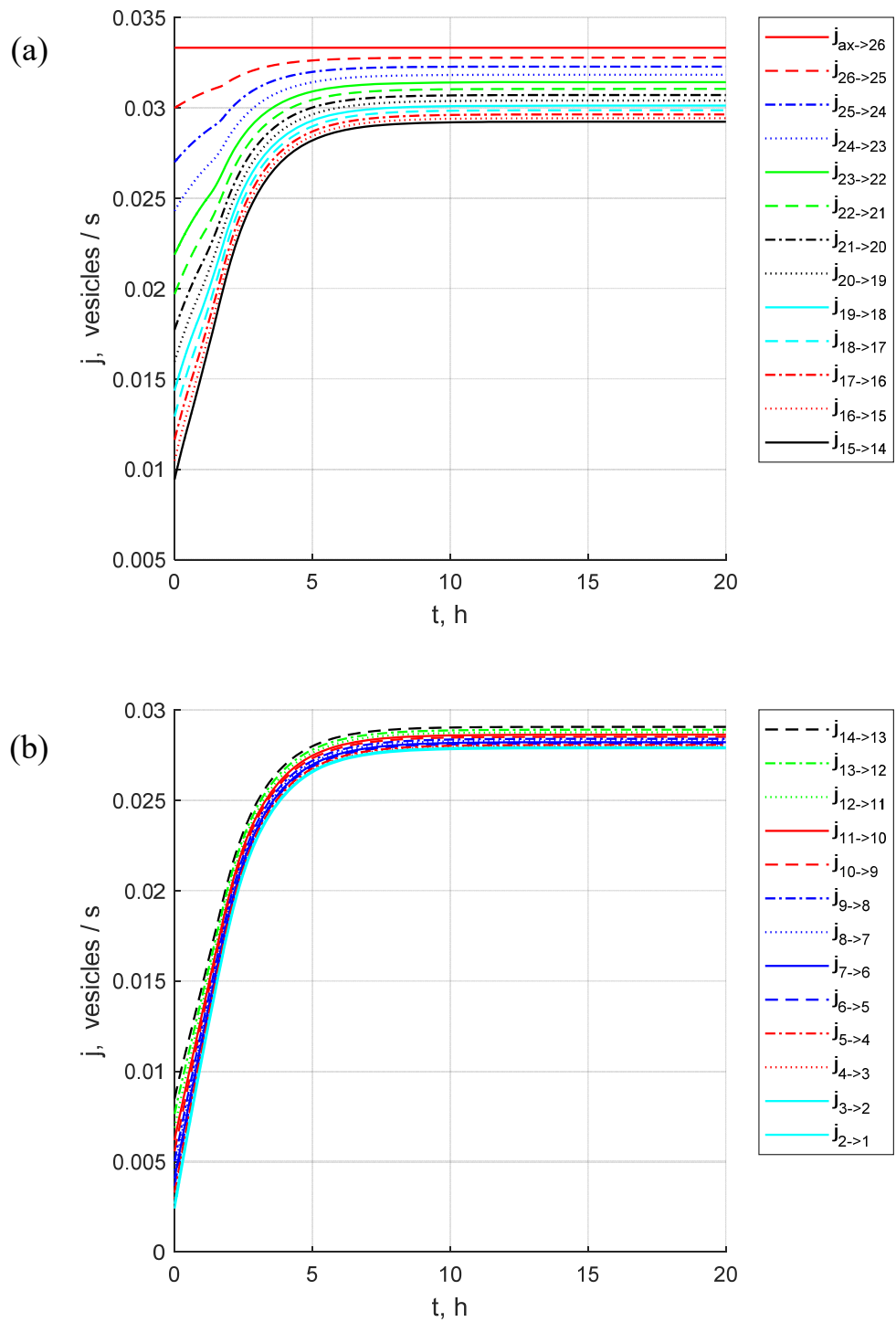

Fig. S4. The buildup toward steady-state: the flux from the axon to the most proximal bouton and anterograde fluxes between various boutons. (a) Fluxes ax→26 through 15→14. (b) Fluxes

14→13 through 2→1. The case when $j_{ax \to 26}$ is kept constant (at 2 DCVs/min), $\delta = 0$ (which refers to the case when all captured DCVs are destroyed in boutons).

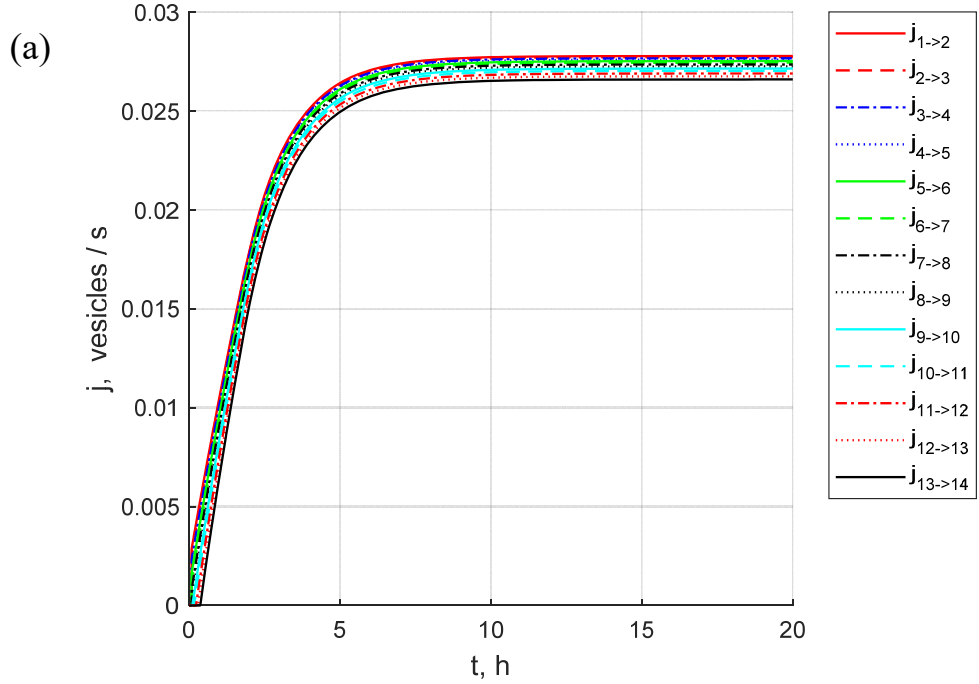

(a)

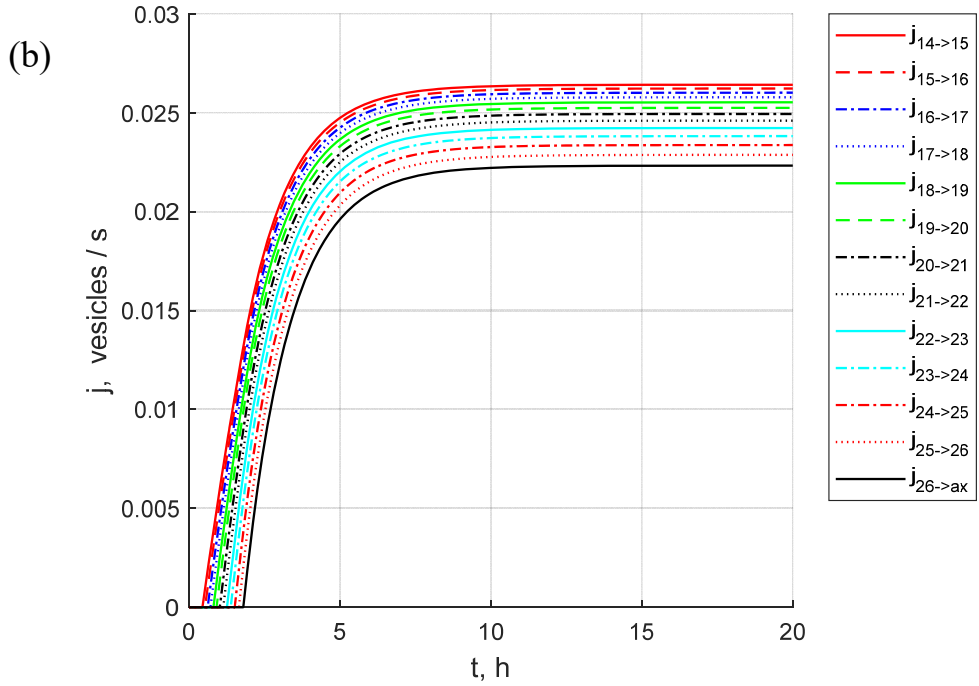

(b)

Fig. S5. The buildup toward steady-state: retrograde fluxes between various boutons and the flux from the most proximal bouton to the axon. (a) Fluxes 1→2 through 13→14. (b) Fluxes 14→15 through 26→ax. The case when $j_{ax \to 26}$ is kept constant (at 2 DCVs/min), $\delta = 0$ (which refers to the case when all captured DCVs are destroyed in boutons).

**S2.4. The case when the DCV flux from the axon to the most proximal bouton is not constant (modeling average DCV concentration in the axon)**

**S2.4.1. DCVs captured into the resident state in boutons escape and re-enter the transiting state ($\delta = 1$)**

**S2.4.1.1. The situation when the half-life of DCVs in the axon, $T_{1/2,ax}$, is equal to the DCV half-residence time in boutons, $T_{1/2}$**

In this series of simulations, the DCV flux from the axon to the most proximal bouton is assumed to be proportional to the average DCV concentration in the axon (Eq. (5)), and the average DCV concentration in the axon is modeled by Eq. (4a). It is assumed that $\delta = 1$ and $T_{1/2,ax} = T_{1/2}$. The results show that in the beginning of filling the terminal ($t$ = 0.2, 1, and 5 h) the DCVs accumulate in more proximal boutons while the distal boutons remain empty (Fig. S6). The decrease in concentration from the most proximal to the most distal boutons is exponential.

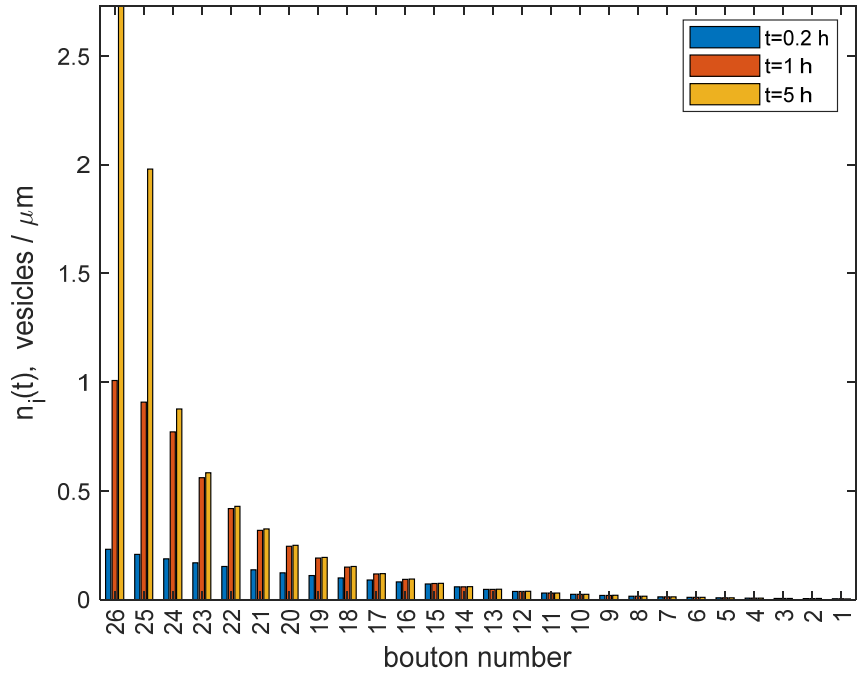

Fig. S6. Concentrations of captured DCVs in various boutons at three times: $t = 0.2$, 1, and 5 h. The case when the average DCV concentration in the axon is modeled by Eq. (4a), and $j_{ax \to 26}$ is modeled by Eq. (5). $\delta = 1$ (which refers to the case when all captured DCVs eventually reenter the transiting pool), $T_{1/2,ax} = 1 \times T_{1/2}$.

It takes approximately 50 hours to fill the boutons to saturation (Fig. S7, Table S3). There are three stages during the process of filling the boutons: 1) the concentrations in the resident state in boutons increase sharply from zero; 2) the bouton concentrations reach their plateaus; 3) concentrations increase again until they reach their steady-state values.

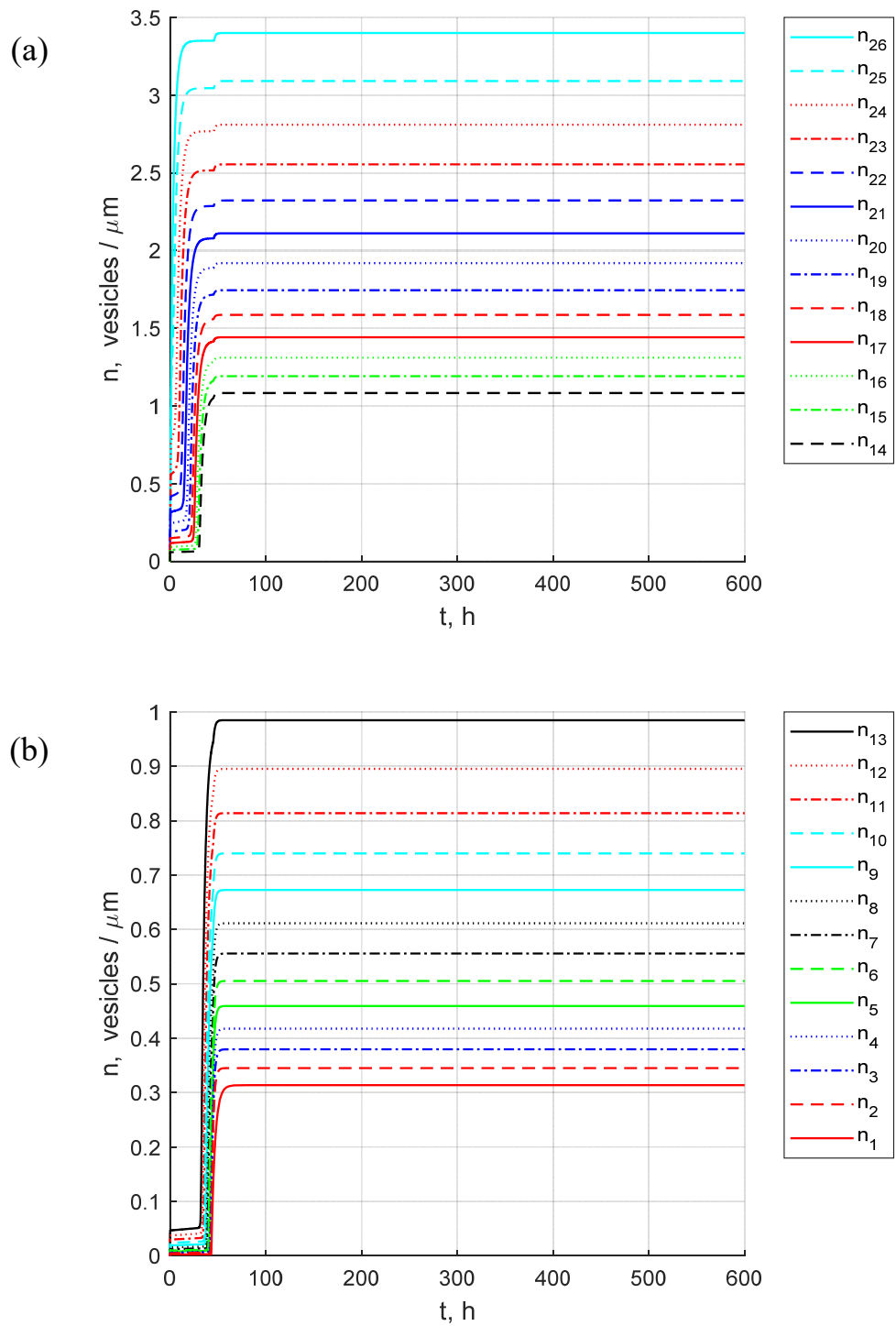

Fig. S7. The buildup toward steady-state: concentrations of captured DCVs in various boutons. (a) Boutons 26 through 14. (b) Boutons 13 through 1. The case when the average DCV

concentration in the axon is modeled by Eq. (4a), and $j_{ax \to 26}$ is modeled by Eq. (5). $\delta = 1$ (which refers to the case when all captured DCVs eventually reenter the transiting pool), $T_{1/2,ax} = 1 \times T_{1/2}$.

The average DCV concentration in the axon first decreases from its initial steady-state value (the last equation in (11)), then reaches a plateau, and then quickly increases until it reaches a steady-state value (Fig. S8). The reason for the initial decrease of the average concentration is the depletion of the axon of DCVs as the terminal is filled. In the beginning of terminal filling, the flux of DCVs into the terminal exceeds the flux of DCVs that enter the axon from the soma. However, when the DCV concentrations in boutons become closer to their saturated values, refilling of the axon begins: the DCV flux into the terminal decreases, and the average DCV concentration in the axon increases back to its steady-state value (the beginning of this process is shown by a rhombus in Fig. S8).

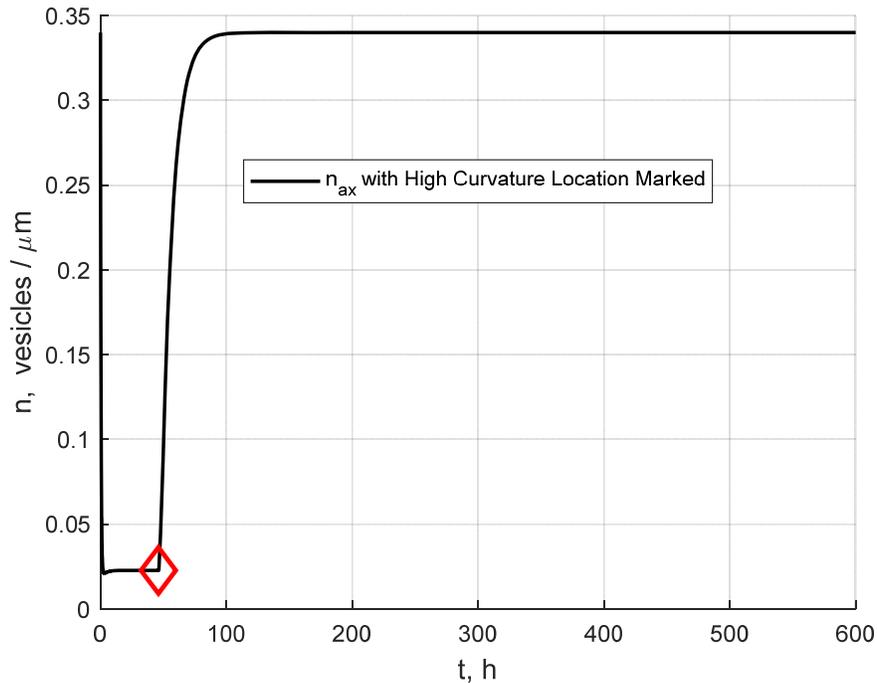

Fig. S8. The buildup toward steady-state: concentrations of transiting DCVs in the axon. A rhombus shows the time of switching between two different transport regimes: (i) the regime when DCV capture in boutons exceeds the rate of DCV synthesis in the soma and (ii) the regime when the terminal is filled to saturation, and the DCV concentration in the axon recovers to its steady-state value. The case when the average DCV concentration in the axon is modeled by Eq. (4a), and $j_{ax \to 26}$ is modeled by Eq. (5). $\delta = 1$ (which refers to the case when all captured DCVs eventually reenter the transiting pool), $T_{1/2,ax} = 1 \times T_{1/2}$.

Anterograde (Fig. S9) and retrograde (Fig. S10) fluxes in the terminal follow the same trend as the DCV concentrations (Fig. S7): they first increase, reach a plateau, and increase again until they reach a steady-state value. Because in this case $\delta = 1$ (no DCV destruction in boutons, all DCVs return to the transiting pool after spending some time in boutons), all DCV fluxes between boutons reach the same steady-state value. This situation thus produces a DCV circulation in the terminal.

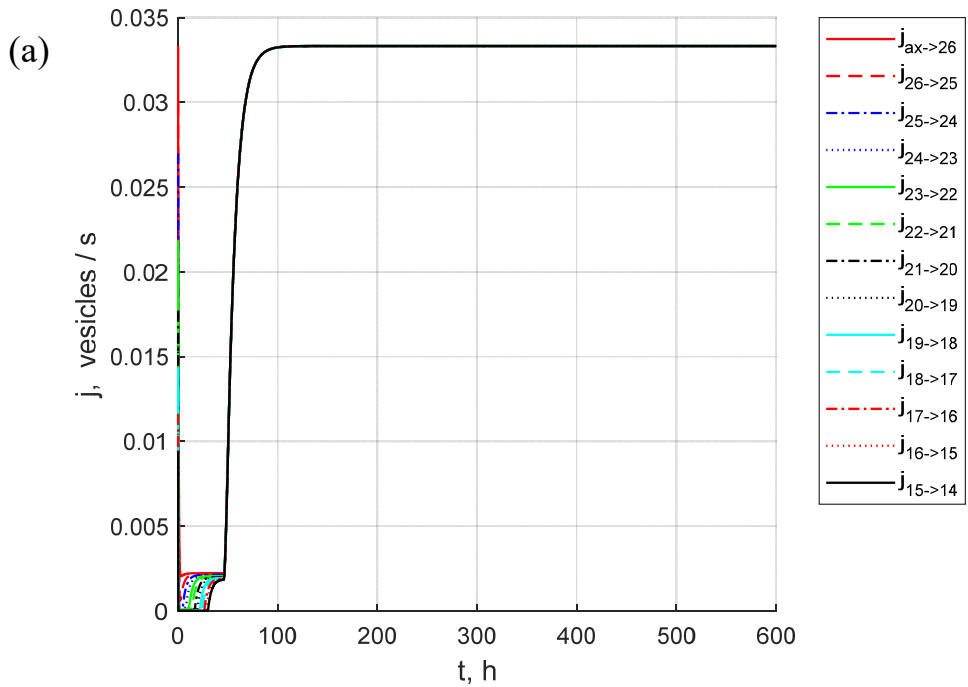

(a)

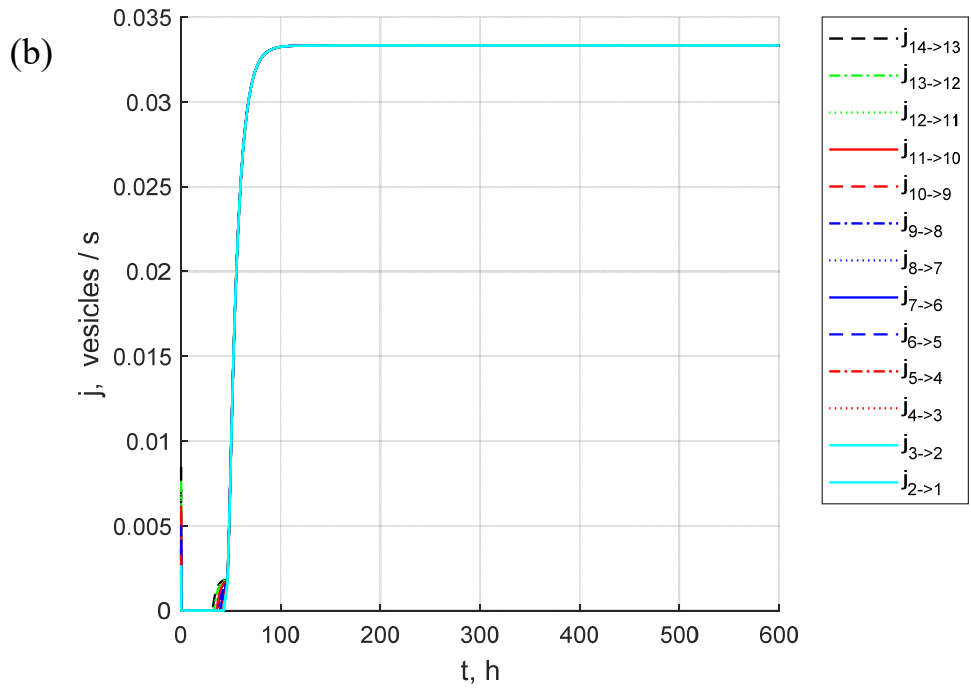

Fig. S9. The buildup toward steady-state: the flux from the axon to the most proximal bouton and anterograde fluxes between various boutons. (a) Fluxes ax→26 through 15→14. (b) Fluxes 14→13 through 2→1. The case when the average DCV concentration in the axon is modeled by Eq. (4a), and $j_{ax \to 26}$ is modeled by Eq. (5). $\delta = 1$ (which refers to the case when all captured DCVs eventually reenter the transiting pool), $T_{1/2,ax} = 1 \times T_{1/2}$.

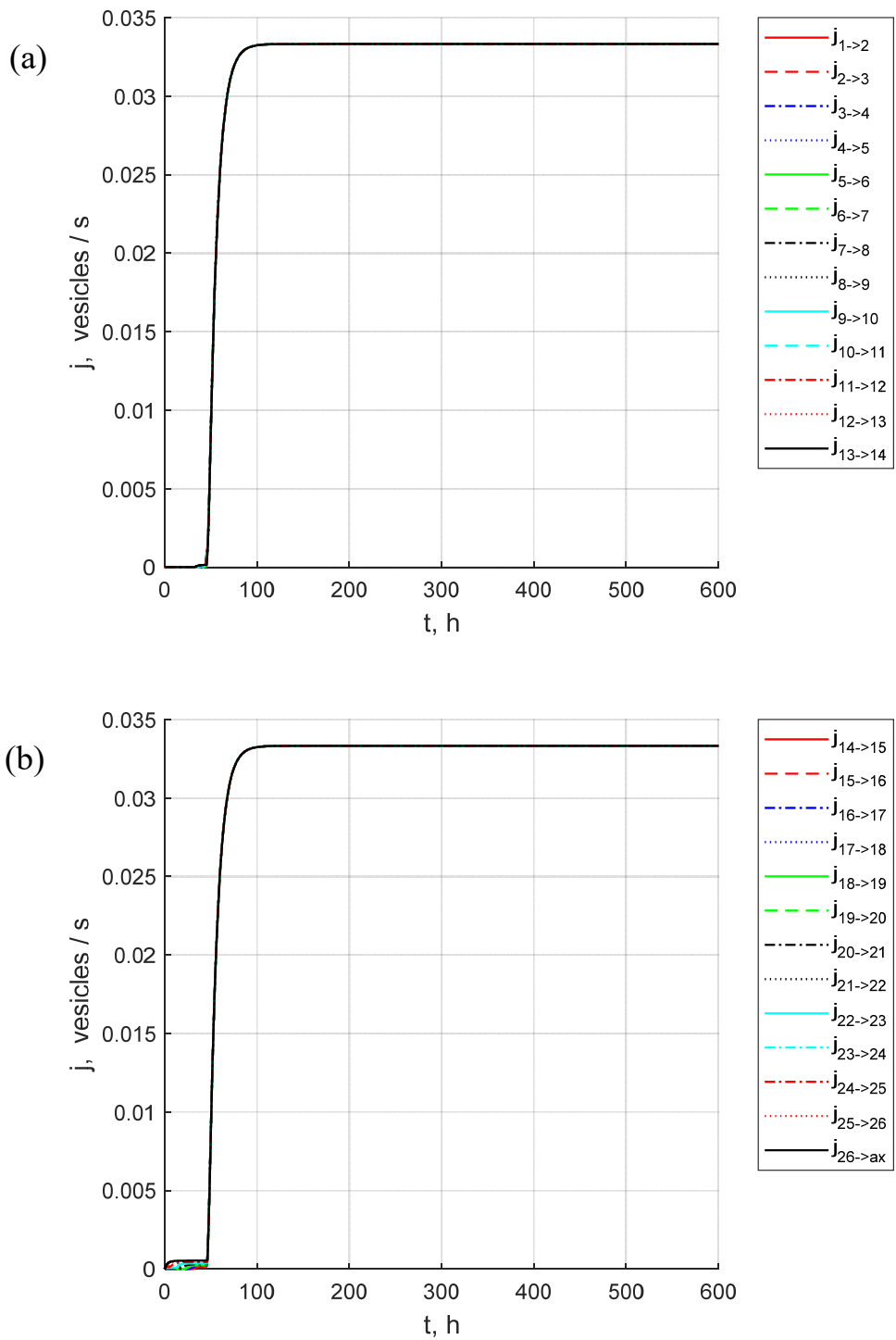

Fig. S10. The buildup toward steady-state: retrograde fluxes between various boutons and the flux from the most proximal bouton to the axon. (a) Fluxes 1→2 through 13→14. (b) Fluxes 14→15 through 26→ax. The case when the average DCV concentration in the axon is modeled

by Eq. (4a), and $j_{ax \to 26}$ is modeled by Eq. (5). $\delta = 1$ (which refers to the case when all captured DCVs eventually reenter the transiting pool), $T_{1/2,ax} = 1 \times T_{1/2}$.

**S2.4.1.2. The situation when the half-life of DCVs in the axon, $T_{1/2,ax}$, is 100 times greater than the DCV half-residence time in boutons, $T_{1/2}$**

If the DCV half-life in the axon is increased by a factor of 100, the time for the DCV concentrations in boutons to reach steady-state increases significantly (to approximately 3,000 hours), as depicted in Fig. S11, see also Table S3 (this obviously cannot be reached in *Drosophila* due to much shorter *Drosophila* life cycle). This is because for a short DCV half-life in the axon, a large number of DCVs are destroyed in the axon, which requires a large supply of DCVs from the soma to compensate for this destruction. However, if the DCV half-life in the axon is large, the required DCV supply from the soma is small, and hence it takes a long time to replenish the initial depletion of DCV concentration in the axon.

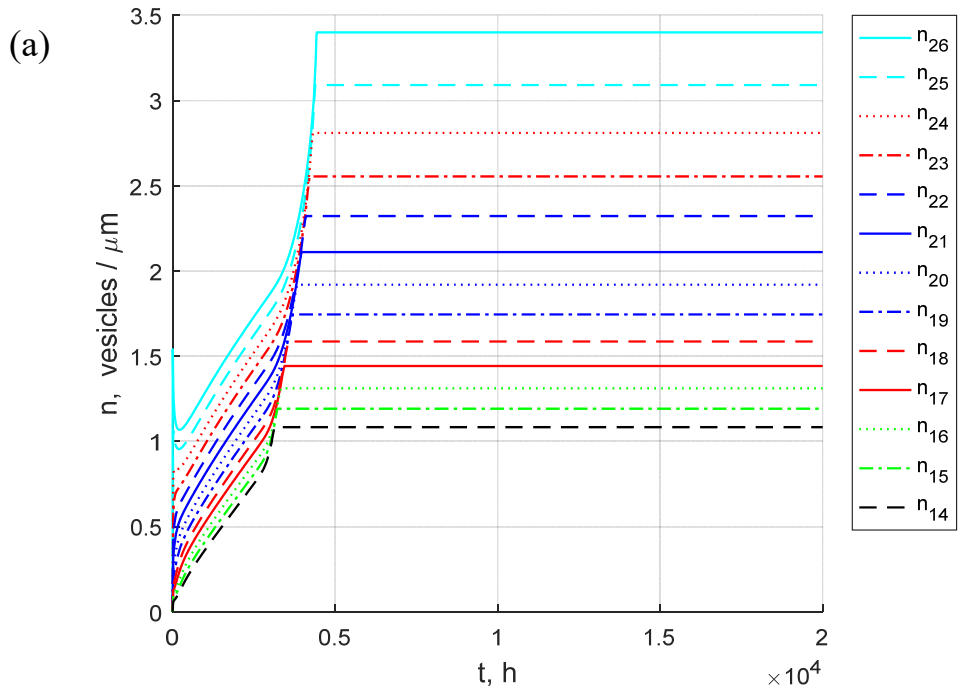

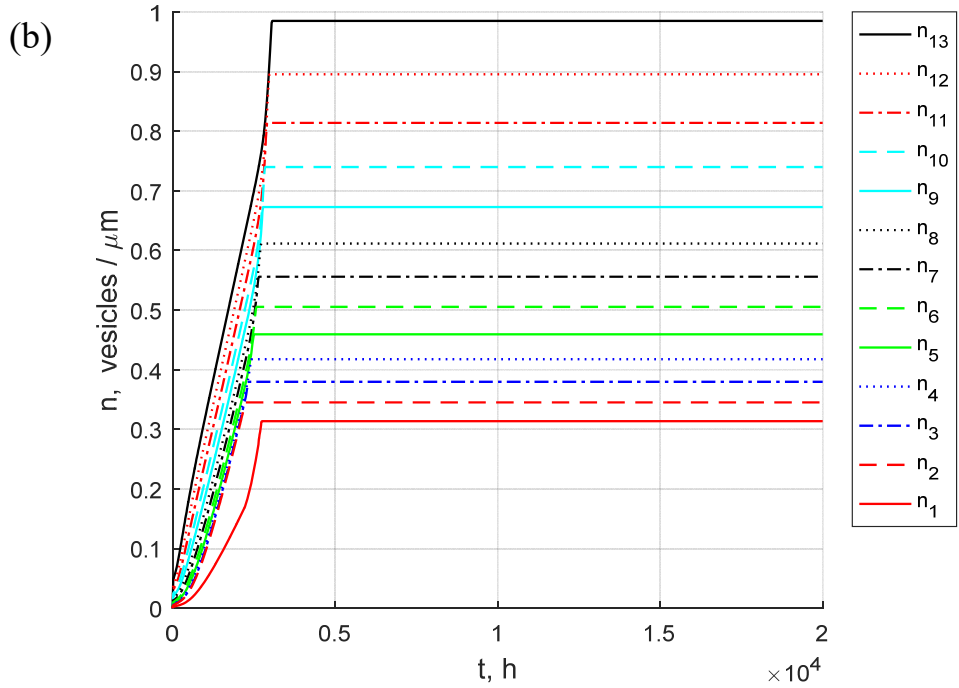

Fig. S11. The buildup toward steady-state: concentrations of captured DCVs in various boutons. (a) Boutons 26 through 14. (b) Boutons 13 through 1. The case when the average DCV concentration in the axon is modeled by Eq. (4a), and $j_{ax \to 26}$ is modeled by Eq. (5). $\delta = 1$ (which refers to the case when all captured DCVs eventually reenter the transiting pool), $T_{1/2,ax} = 100 \times T_{1/2}$.

Qualitatively, the average DCV concentration in the axon behaves similarly to that for $T_{1/2,ax} = T_{1/2}$, but it now takes much longer, approximately 4,400 hours (Fig. S12, Table S6), for the DCV concentration in the axon to recover from the drop caused by the depletion of the axon of DCVs due to filling the terminal and begin to increase back to its steady-state value (the time of 4,400 hours is certainly impossible in *Drosophila*).

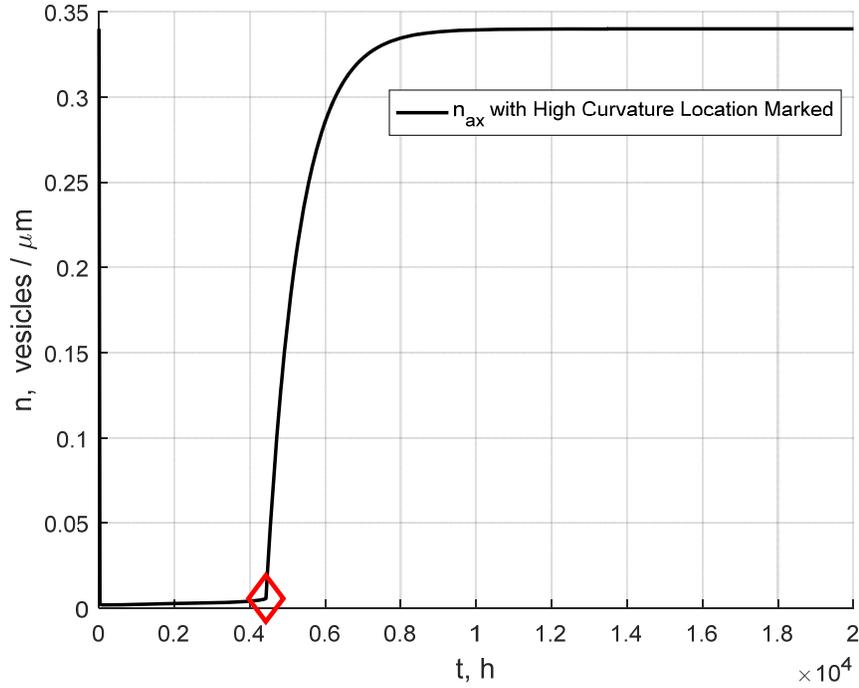

Fig. S12. The buildup toward steady-state: concentrations of transiting DCVs in the axon. A rhombus shows the time of switching between two different transport regimes: (1) the regime when DCV capture in boutons exceeds the rate of DCV synthesis in the soma and (2) the regime when the terminal is filled to saturation, and the DCV concentration in the axon recovers to its steady-state value. The case when the average DCV concentration in the axon is modeled by Eq. (4a), and $j_{ax \to 26}$ is modeled by Eq. (5). $\delta = 1$ (which refers to the case when all captured DCVs eventually reenter the transiting pool), $T_{1/2,ax} = 100 \times T_{1/2}$.

Anterograde (Fig. S13) and retrograde (Fig. S14) fluxes are very similar in all boutons and exhibit the same behavior: a rapid decrease, an almost constant value, and a slow increase to steady-state. Both anterograde and retrograde fluxes reach their steady-state values in approximately 8,000 hours (Table S4).

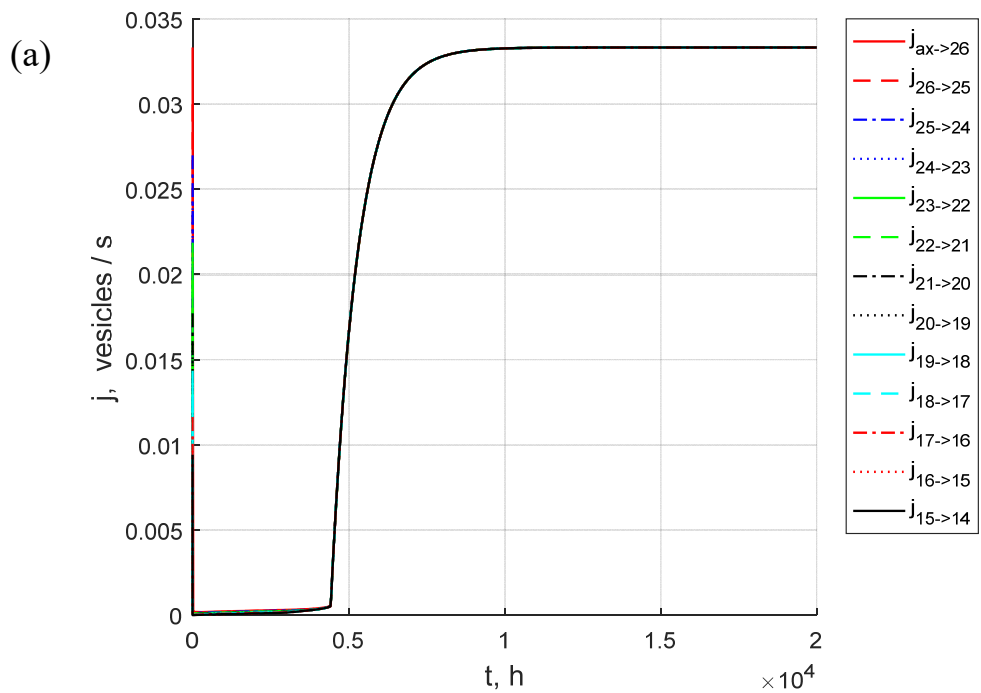

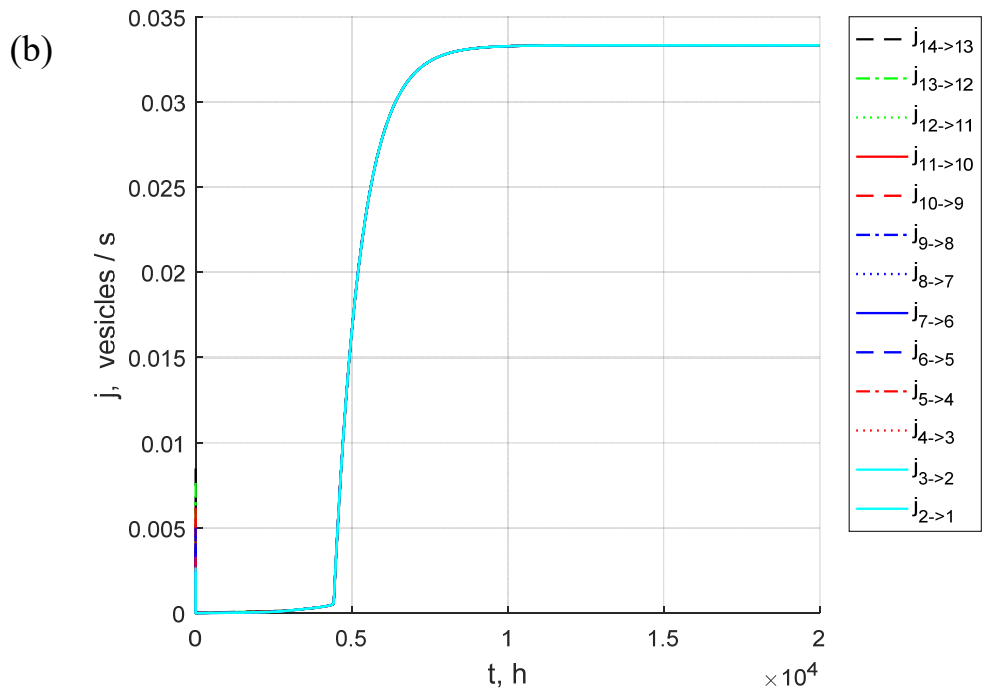

Fig. S13. The buildup toward steady-state: the flux from the axon to the most proximal bouton and anterograde fluxes between various boutons. (a) Fluxes ax→26 through 15→14. (b) Fluxes 14→13 through 2→1. The case when the average DCV concentration in the axon is modeled by Eq. (4a), and $j_{ax \to 26}$ is modeled by Eq. (5). $\delta = 1$ (which refers to the case when all captured DCVs eventually reenter the transiting pool), $T_{1/2,ax} = 100 \times T_{1/2}$.

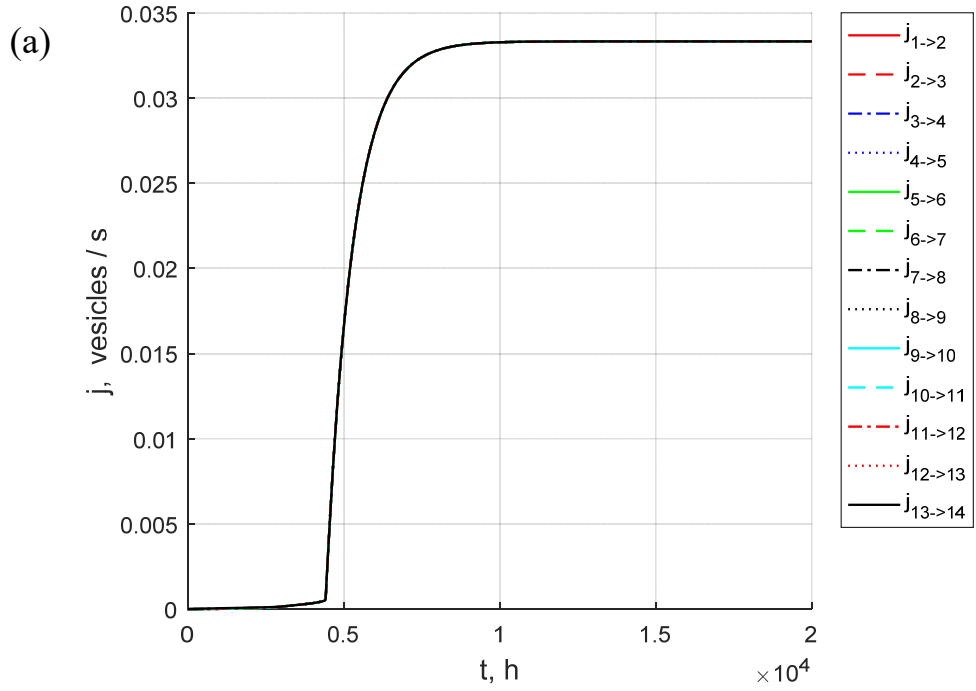

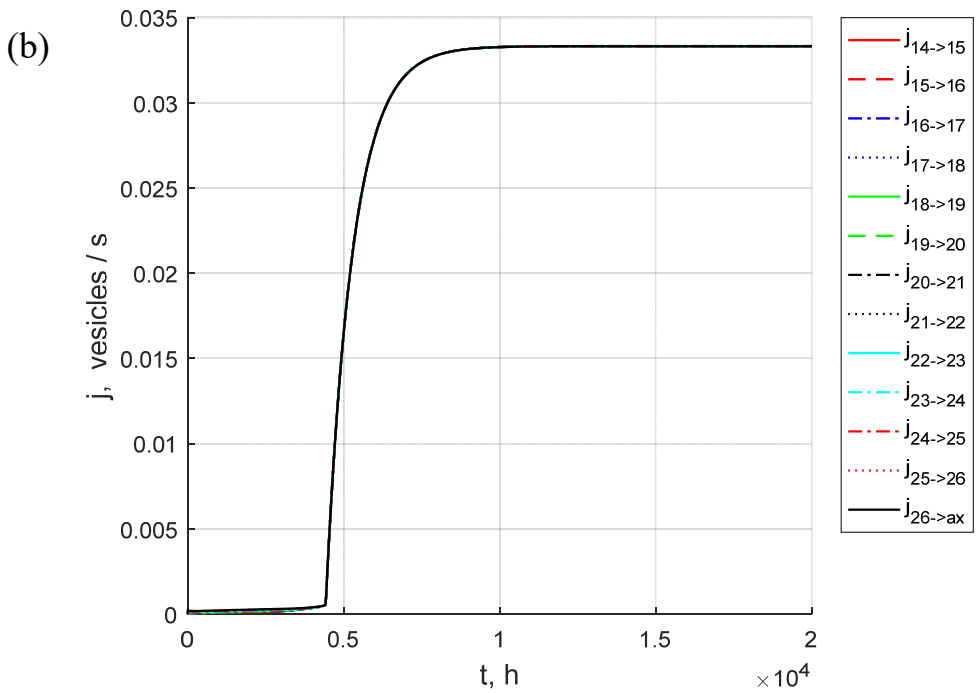

Fig. S14. The buildup toward steady-state: retrograde fluxes between various boutons and the flux from the most proximal bouton to the axon. (a) Fluxes 1→2 through 13→14. (b) Fluxes 14→15 through 26→ax. The case when the average DCV concentration in the axon is modeled by Eq. (4a), and $j_{ax \to 26}$ is modeled by Eq. (5). $\delta = 1$ (which refers to the case when all captured DCVs eventually reenter the transiting pool), $T_{1/2,ax} = 100 \times T_{1/2}$.

### S2.4.2. DCVs captured into the resident state are destroyed in boutons ($\delta = 0$)

### S2.4.2.1. The situation when the half-life of DCVs in the axon, $T_{1/2,ax}$, is equal to the DCV half-residence time in boutons, $T_{1/2}$

In this case, at relatively small times ($t$ = 0.2, 1, and 5 h), the proximal half of the terminal accumulates DCVs while the distal half remains empty (Fig. S15).

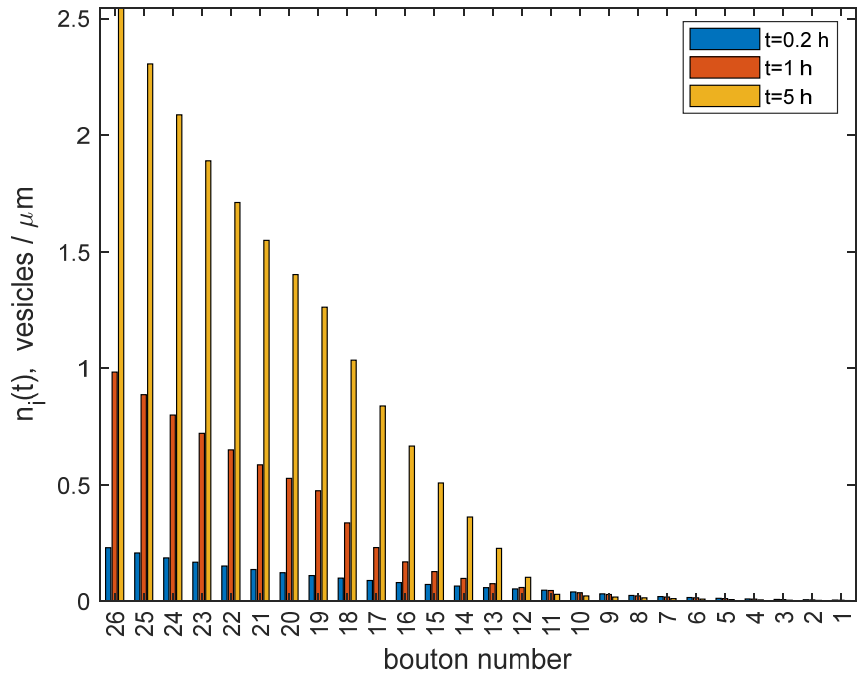

Fig. S15. Concentrations of captured DCVs in various boutons at three times: $t$ = 0.2, 1, and 5 h. The case when the average DCV concentration in the axon is modeled by Eq. (4a), and $j_{ax \to 26}$ is modeled by Eq. (5). $\delta = 0$ (which refers to the case when all captured DCVs are destroyed in boutons), $T_{1/2,ax} = 1 \times T_{1/2}$.

Interestingly, for $\delta = 0$, when DCVs are destroyed in boutons, the boutons are filled quicker than for the similar case of $\delta = 1$ (compare Figs. S16 and S7). It now takes approximately 20 hours to fill the boutons to saturation (Fig. S16, Table S3). This is because for $\delta = 0$ the model assumes that DCVs are destroyed in the terminal, which leads to more DCV flux from the soma into the axon (we assumed that enough DCVs are produced in the soma to compensate for the DCV destruction in the axon and soma at steady-state).

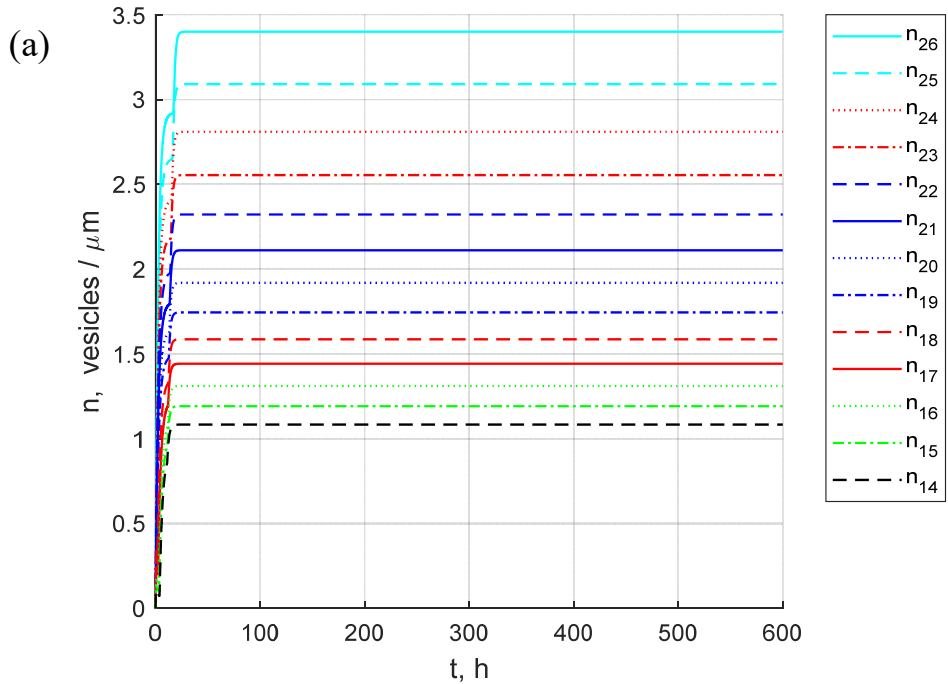

(a)

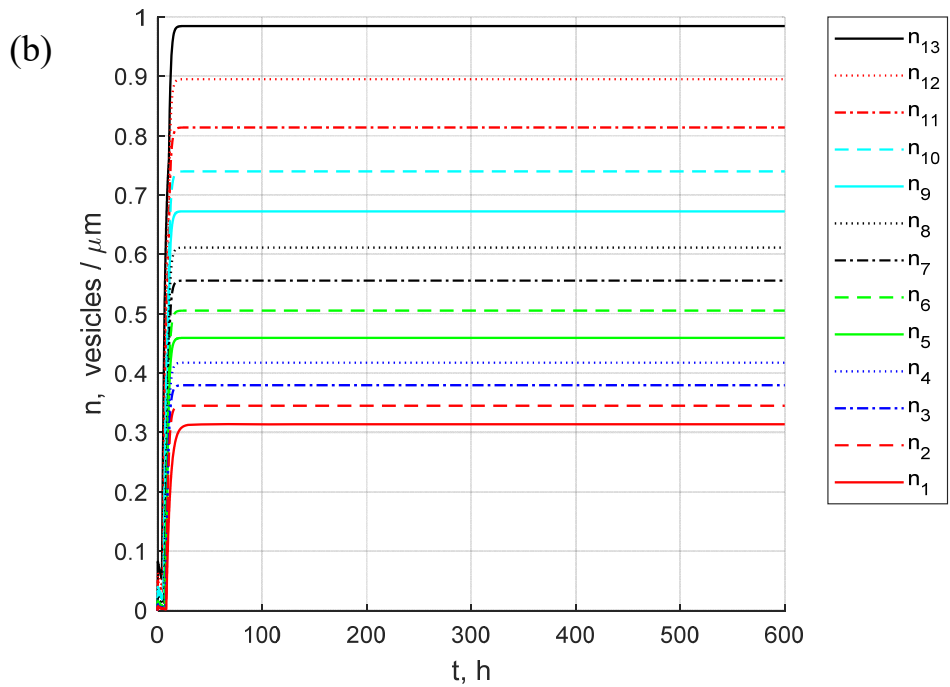

Fig. S16. The buildup toward steady-state: concentrations of captured DCVs in various boutons. (a) Boutons 26 through 14. (b) Boutons 13 through 1. The case when the average DCV concentration in the axon is modeled by Eq. (4a), and $j_{ax \to 26}$ is modeled by Eq. (5). $\delta = 0$ (which refers to the case when all captured DCVs are destroyed in boutons), $T_{1/2,ax} = 1 \times T_{1/2}$.

It takes approximately 20 hours for the axonal DCV concentration to reach the end of the plateau region (see the position of the rhombus in Fig. S17, Table S6), which is significantly shorter than that for $\delta = 1$ (Fig. S8, Table S6).

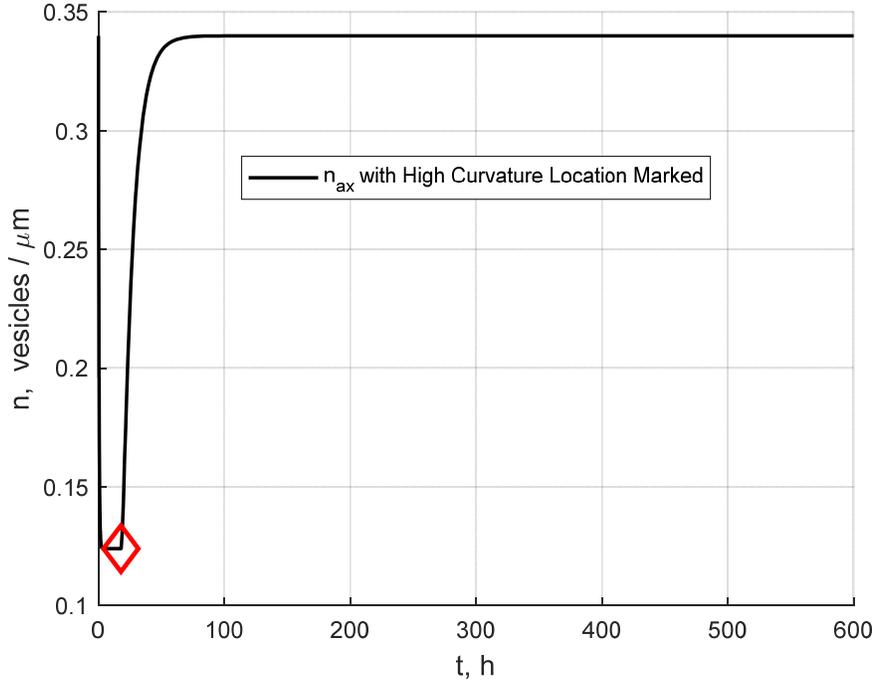

Fig. S17. The buildup toward steady-state: concentrations of transiting DCVs in the axon. A rhombus shows the time of switching between two different transport regimes: (1) the regime when DCV capture in boutons exceeds the rate of DCV synthesis in the soma and (2) the regime when the terminal is filled to saturation, and the DCV concentration in the axon recovers to its steady-state value. The case when the average DCV concentration in the axon is modeled by Eq. (4a), and $j_{ax \to 26}$ is modeled by Eq. (5). $\delta = 0$ (which refers to the case when all captured DCVs are destroyed in boutons), $T_{1/2,ax} = 1 \times T_{1/2}$.

Anterograde (Fig. S18) and retrograde (Fig. S19) fluxes increase to slightly different values in different boutons within approximately 60 hours (Table S4).

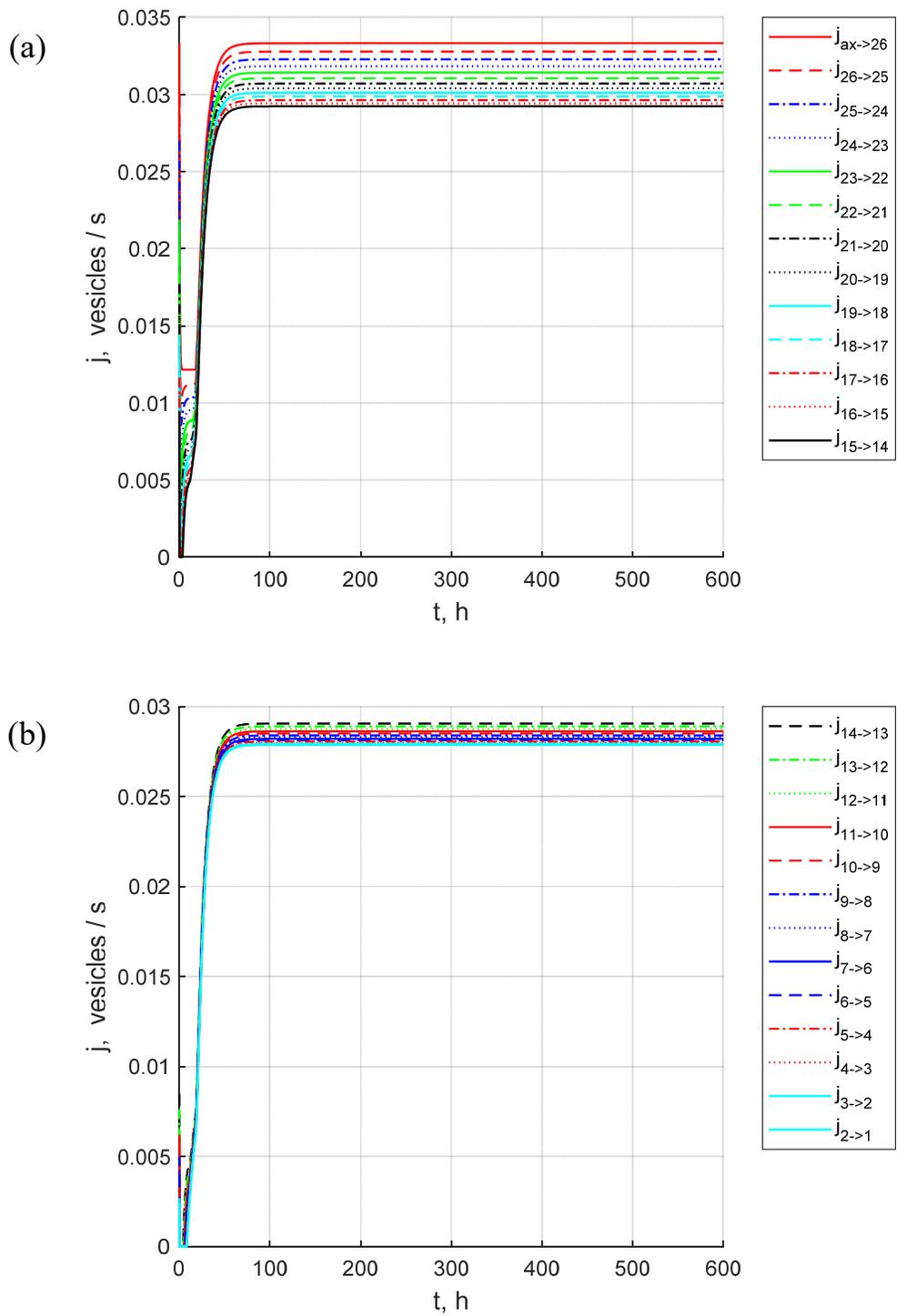

Fig. S18. The buildup toward steady-state: the flux from the axon to the most proximal bouton and anterograde fluxes between various boutons. (a) Fluxes ax→26 through 15→14. (b) Fluxes 14→13 through 2→1. The case when the average DCV concentration in the axon is modeled by Eq. (4a), and $j_{ax \to 26}$ is modeled by Eq. (5). $\delta = 0$ (which refers to the case when all captured DCVs are destroyed in boutons), $T_{1/2,ax} = 1 \times T_{1/2}$.

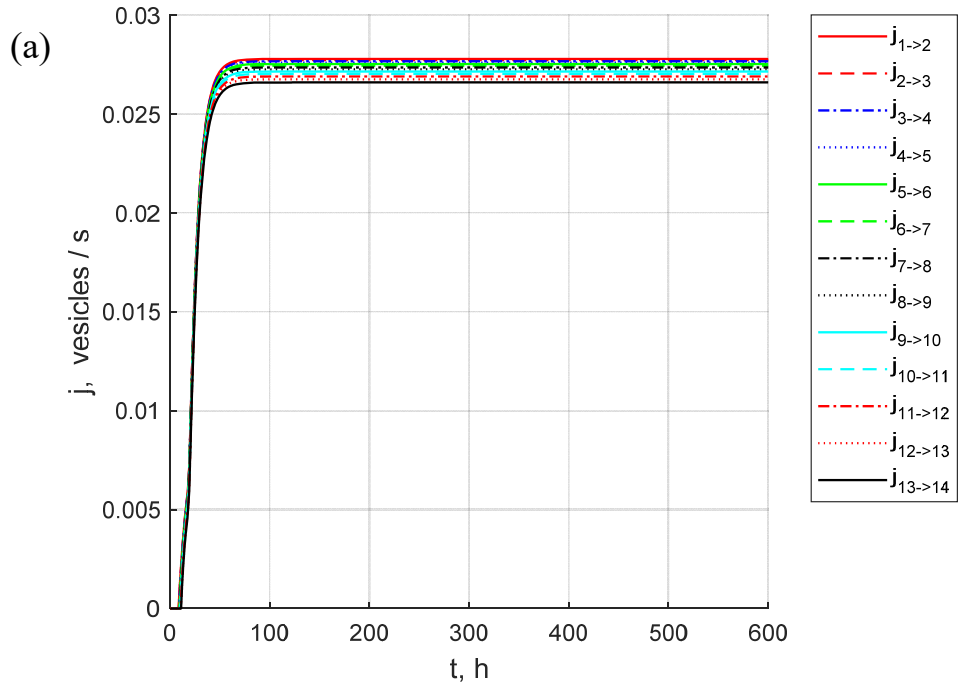

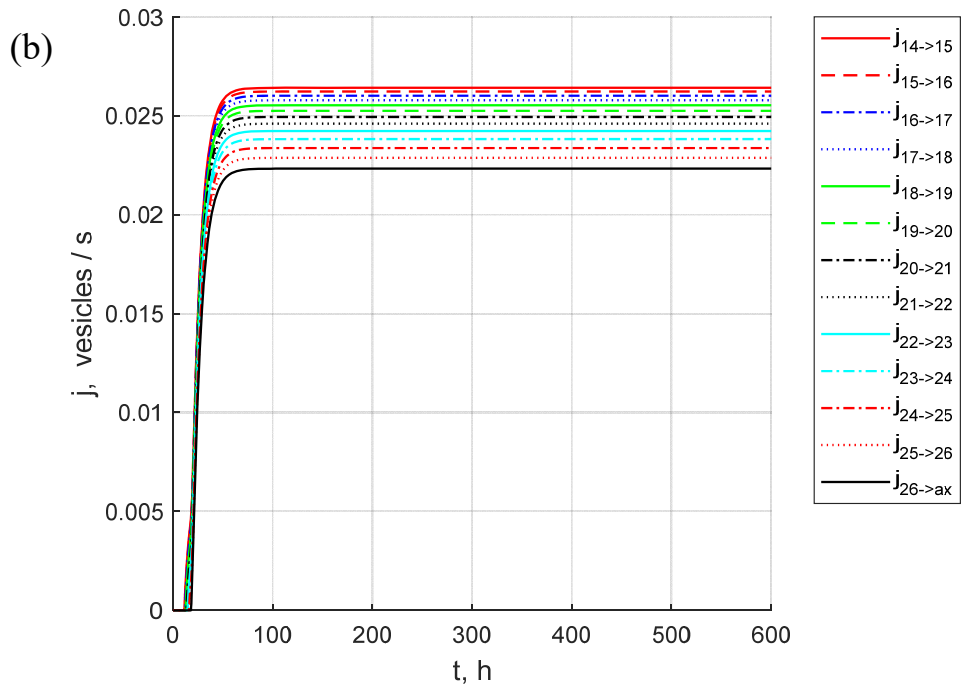

Fig. S19. The buildup toward steady-state: retrograde fluxes between various boutons and the flux from the most proximal bouton to the axon. (a) Fluxes 1→2 through 13→14. (b) Fluxes 14→15 through 26→ax. The case when the average DCV concentration in the axon is modeled by Eq. (4a), and $j_{ax \to 26}$ is modeled by Eq. (5). $\delta = 0$ (which refers to the case when all captured DCVs are destroyed in boutons), $T_{1/2,ax} = 1 \times T_{1/2}$.

## S2.4.2.2. The situation when the half-life of DCVs in the axon, $T_{1/2,ax}$, is 100 times greater than the DCV half-residence time in boutons, $T_{1/2}$

It is interesting that if the half-life of DCVs in the axon is increased ($T_{1/2,ax} = 100 \times T_{1/2}$), the time it takes for the DCV concentrations in boutons to reach their saturated values (between 20 and 50 hours depending on the position of the bouton, see Table S3) is much shorter than the time of recovery of the axonal concentration to its steady-state value (compare Figs. S20 and S21, see also Table S3). This is because in this case, the concentration in the axon does not decrease to a very low value (Fig. S21). Since the DCV flux into the axon is proportional to the axonal concentration of DCVs (Eq. (5)), the terminal in this case fills up relatively quickly.

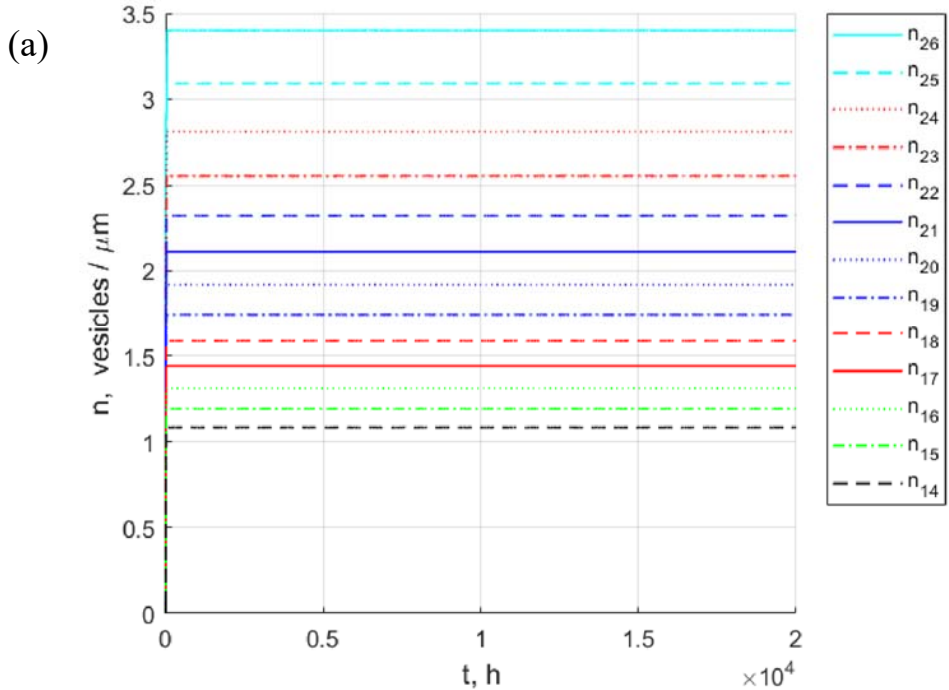

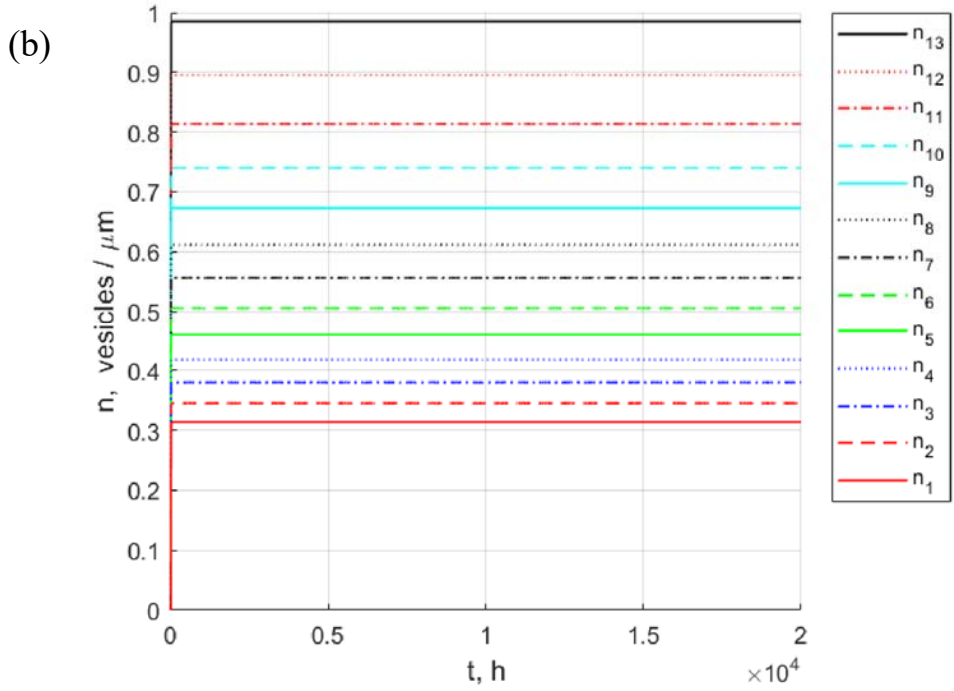

Fig. S20. The buildup toward steady-state: concentrations of captured DCVs in various boutons. (a) Boutons 26 through 14. (b) Boutons 13 through 1. The case when the average DCV concentration in the axon is modeled by Eq. (4a), and $j_{ax \to 26}$ is modeled by Eq. (5). $\delta = 0$ (which refers to the case when all captured DCVs are destroyed in boutons), $T_{1/2,ax} = 100 \times T_{1/2}$.

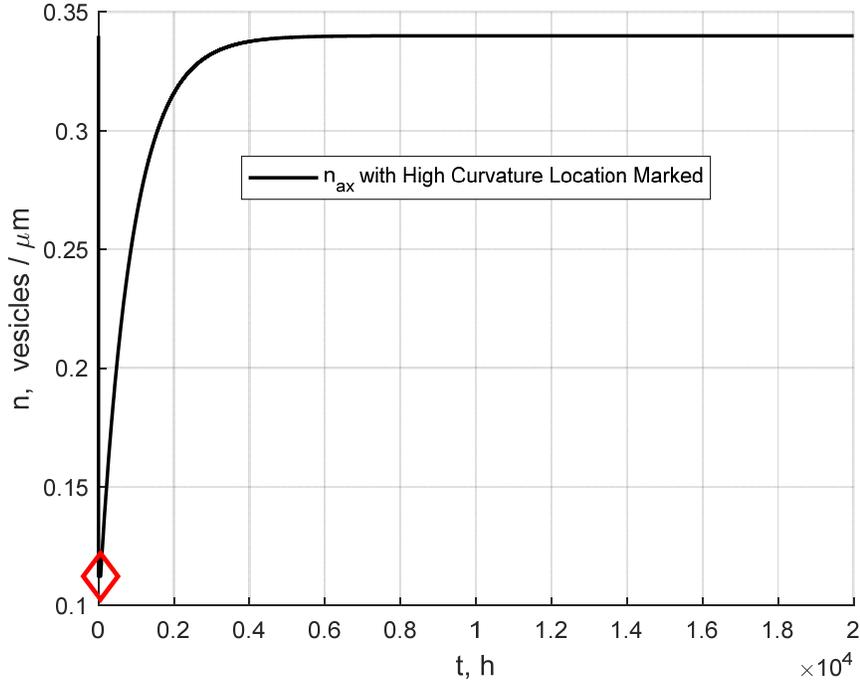

Fig. S21. The buildup toward steady-state: concentrations of transiting DCVs in the axon. A rhombus shows the time of switching between two different transport regimes: (1) the regime when DCV capture in boutons exceeds the rate of DCV synthesis in the soma and (2) the regime when the terminal is filled to saturation, and the DCV concentration in the axon recovers to its steady-state value. The case when the average DCV concentration in the axon is modeled by Eq. (4a), and $j_{ax \to 26}$ is modeled by Eq. (5). $\delta = 0$ (which refers to the case when all captured DCVs are destroyed in boutons), $T_{1/2,ax} = 100 \times T_{1/2}$.

It takes anterograde (Fig. S22) and retrograde (Fig. S23) fluxes between the boutons approximately 4,000 hours to reach their steady-state values (Table S4).

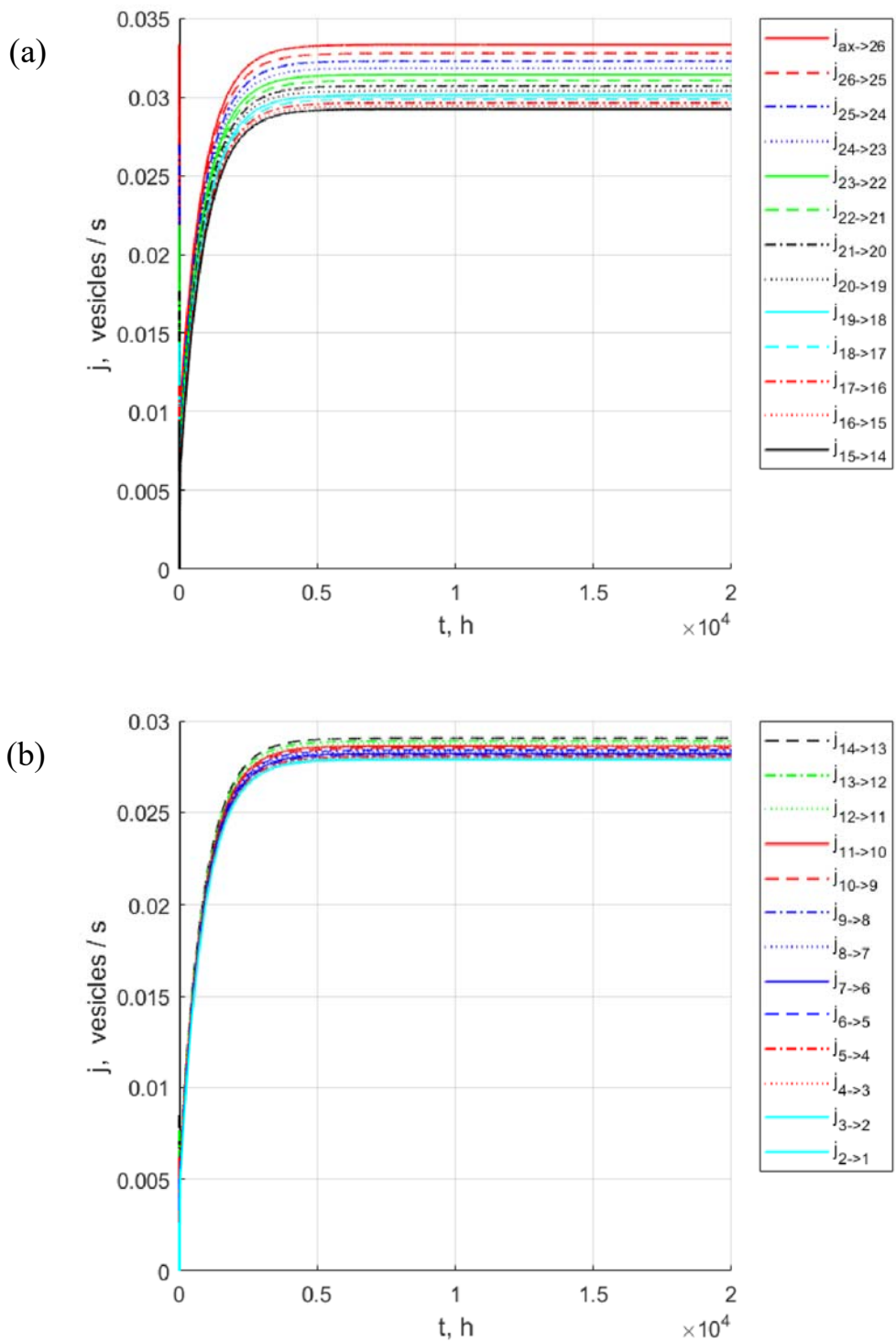

Fig. S22. The buildup toward steady-state: the flux from the axon to the most proximal bouton and anterograde fluxes between various boutons. (a) Fluxes ax→26 through 15→14. (b) Fluxes 14→13 through 2→1. The case when the average DCV concentration in the axon is modeled by Eq. (4a), and $j_{ax \to 26}$ is modeled by Eq. (5). $\delta = 0$ (which refers to the case when all captured DCVs are destroyed in boutons), $T_{1/2,ax} = 100 \times T_{1/2}$.

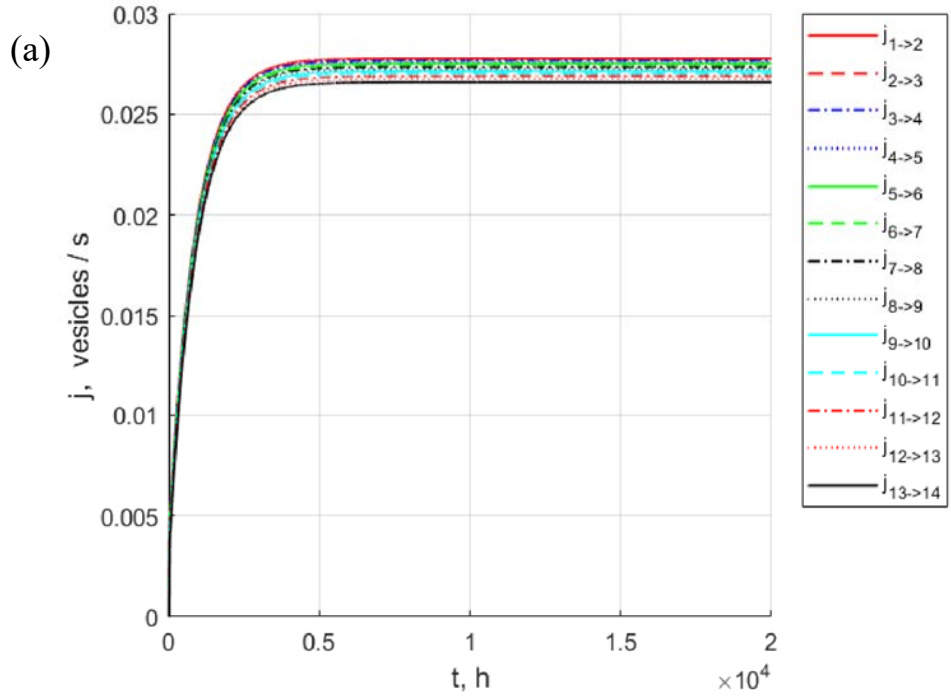

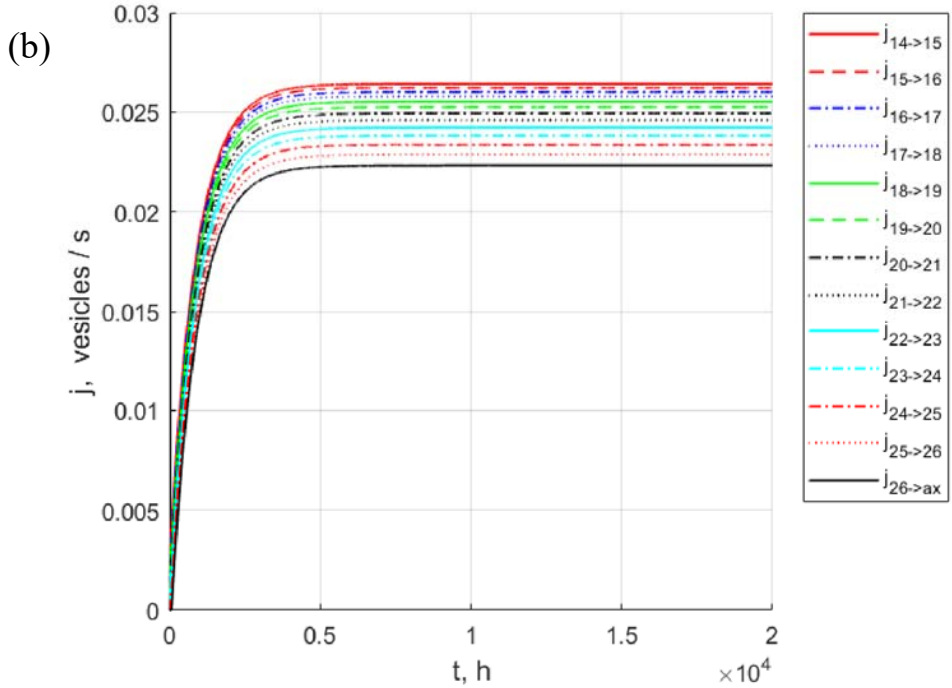

Fig. S23. The buildup toward steady-state: retrograde fluxes between various boutons and the flux from the most proximal bouton to the axon. (a) Fluxes 1→2 through 13→14. (b) Fluxes 14→15 through 26→ax. The case when the average DCV concentration in the axon is modeled by Eq. (4a), and $j_{ax \to 26}$ is modeled by Eq. (5). $\delta = 0$ (which refers to the case when all captured DCVs are destroyed in boutons), $T_{1/2,ax} = 100 \times T_{1/2}$.

## S2.5. Analysis of the dependence of the time of switching between different transport regimes on the fates of DCVs in boutons and on the half-life of DCVs in the axon

The time of switching between different transport regimes, $t_{switch}$ (see the position of the rhombus in Figs. S8, S12, S17, and S21), shows the moment when the axon starts to get refilled after the initial depletion of DCVs, and the DCV concentration in the axon starts to increase. The increase of the DCV half-life in the axon increases the duration during which the axon remains depleted of DCVs (Fig. S24a). This is because when the DCV half-life in the axon is small, a large portion of DCVs are destroyed in the axon before reaching the terminal. To make a steady-state possible, this requires DCVs to be synthesized in the soma at a greater rate. A large supply of DCVs from the soma allows the axon to recover its DCV concentration quickly after filling the terminal.

A sharp increase of $t_{switch}$ with the increase of $\delta$ (Fig. S24b) is explained by the fact that for a greater value of $\delta$, a larger portion of DCVs are returned back to the transiting pool (rather than being destroyed in boutons). This means that for greater $\delta$, less DCV synthesis in the soma is needed to compensate for DCV destruction. A lesser supply of DCVs leads to a longer time that it takes to replenish the initial depletion of the axon of DCVs that happens as DCVs are filling the terminal that is assumed not to contain any DCVs initially (Eq. (11)).

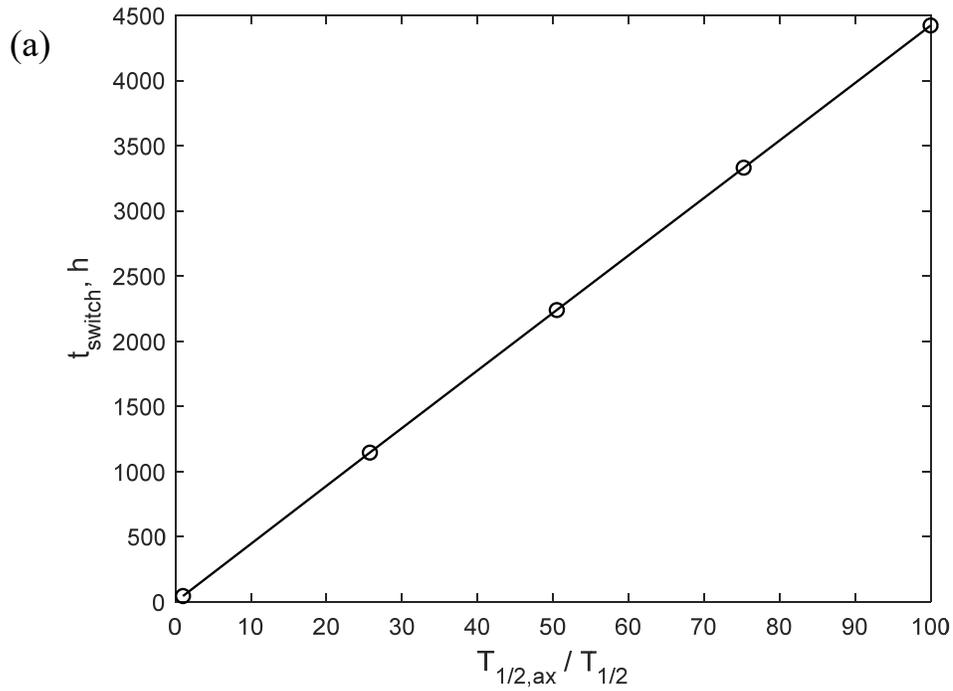

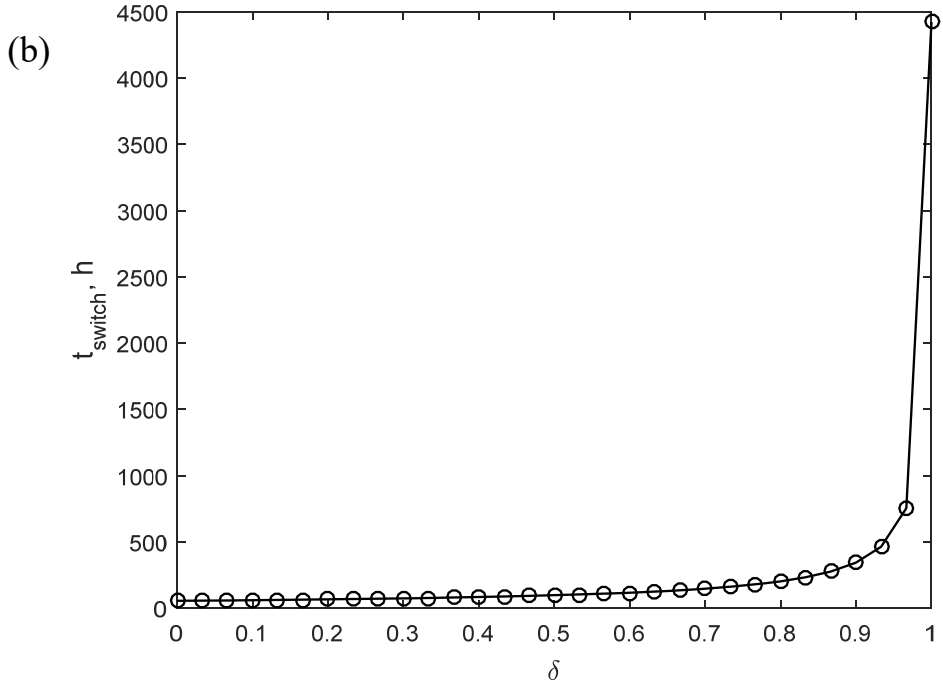

Fig. S24. Time of switching between different transport regimes versus (a) The ratio of the DCV half-life in the axon to DCV half-residence time in boutons, for $\delta = 1$; (b) parameter $\delta$ simulating the fate of resident DCVs ($\delta = 0$ simulates the situation when all captured DCVs are eventually destroyed in boutons and $\delta = 1$ simulates the situation when all captured DCVs eventually reenter the transiting pool), for $T_{1/2,ax} = 100 \times T_{1/2}$.

## S2.6. Age distribution of DCVs in boutons and mean age of DCVs in boutons: Analysis for the case when governing equations (1)-(3) can be linearized

From the numerical solution, we concluded that the anterograde fluxes remain positive at all times (Fig. 6) while the retrograde fluxes become positive in less than one hour into the process of filling the terminal (Fig. 7). Thus, the following approximations can be made:

$$\min\left[ h_{26}^a \left( n_{sat0,26} - n_{26} \right), j_{ax \to 26} \right] = h_{26}^a \left( n_{sat0,26} - n_{26} \right) \quad \text{(S1)}$$

and

$$\min\left[ h_{26}^r \left( n_{sat0,26} - n_{26} \right), j_{25 \to 26} \right] = h_{26}^r \left( n_{sat0,26} - n_{26} \right). \quad \text{(S2)}$$

Eq. (1) can then be simplified as:

$$L_{26}\frac{dn_{26}}{dt} = -\left[h_{26}^a + h_{26}^r + L_{26}\frac{\ln(2)}{T_{1/2}}\right]n_{26} + \left(h_{26}^a + h_{26}^r\right)n_{sat0,26}. \tag{S3}$$

Using the same approximation, Eq. (2) can be simplified as:

$$L_i\frac{dn_i}{dt} = -\left[h_i^a + h_i^r + L_i\frac{\ln(2)}{T_{1/2}}\right]n_i + \left(h_i^a + h_i^r\right)n_{sat0,i} \qquad (i=25,24,\ldots,2), \tag{S4}$$

and Eq. (3) further reduces to:

$$L_1\frac{dn_1}{dt} = -\left[h_1 + L_1\frac{\ln(2)}{T_{1/2}}\right]n_1 + h_1 n_{sat0,1}. \tag{S5}$$

This means that we can linearize Eq. (21). Matrix B for the compartmental system displayed in Fig. 2 is a diagonal matrix whose elements on the main diagonal are given by Eq. (22).

The last element of vector **u** now is

$$u_{26} = \left(h_{26}^a + h_{26}^r\right)\left(n_{sat0,26} - n_{26}\right)/L_{26}. \tag{S6}$$

The other elements of vector **u** now are

$$u_i = \left(h_i^a + h_i^r\right)\left(n_{sat0,i} - n_i\right)/L_i \qquad (i=25,24,\ldots,2) \tag{S7}$$

and

$$u_1 = h_1\left(n_{sat0,1} - n_1\right)/L_1. \tag{S8}$$

We then followed the procedure outlined by Eqs. (26)-(32) and obtained the results displayed in Fig. S25, which are practically identical to the results presented in Fig. 8 for the non-linear case.

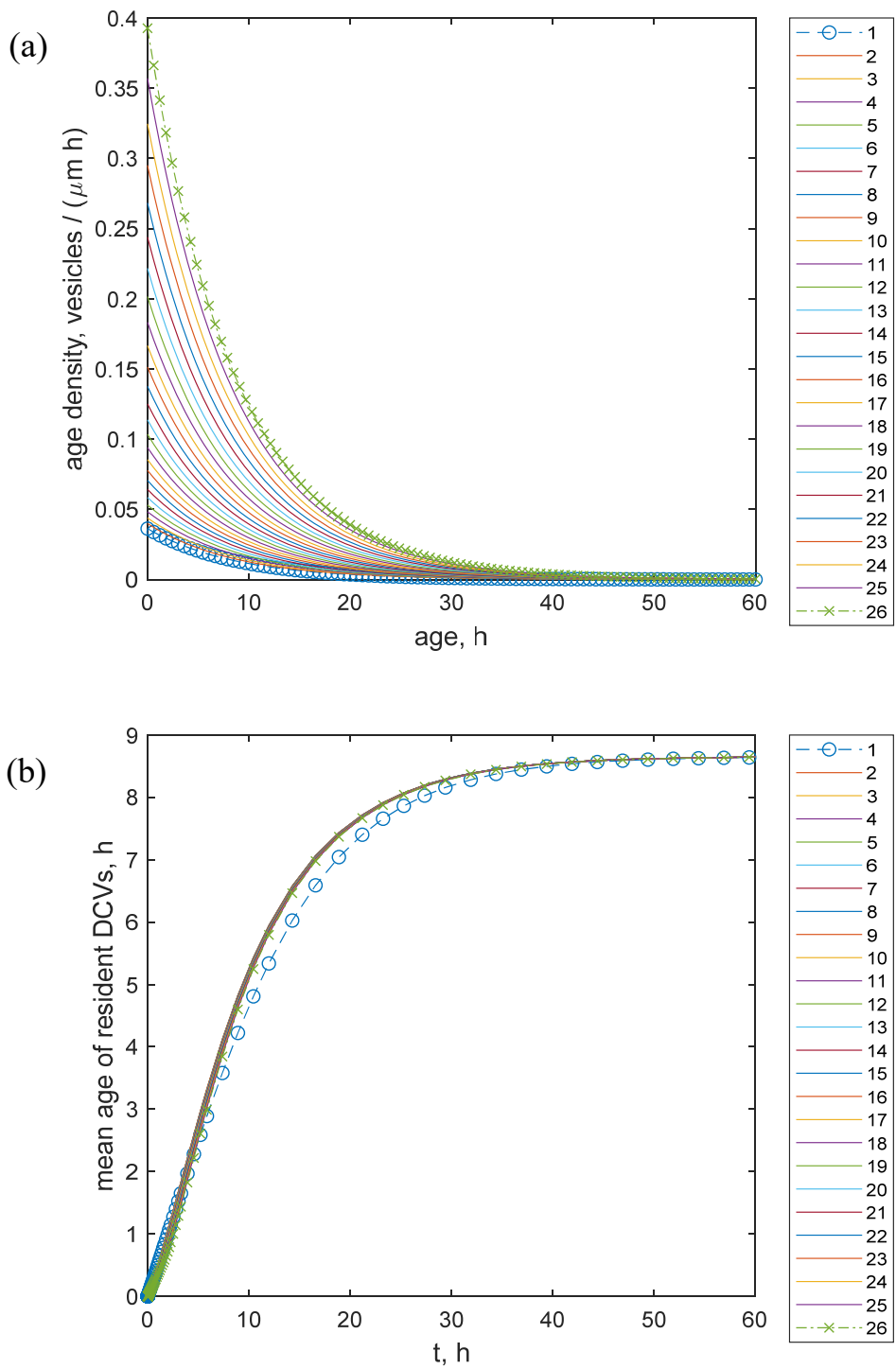

Fig. S25. The case of linearized governing equations: Eqs. (1)-(3) are replaced with their linearized versions given by Eqs. (S3)-(S5). (a) Age density of DCVs in various boutons at steady-state. (b) Mean age of resident DCVs in various boutons versus time. The case when $j_{ax \to 26}$ is kept constant (at 2 DCVs/min). Also note the similarity with Fig. 8.